\documentclass[12pt] {iopart}

\expandafter\let\csname equation*\endcsname\relax
\expandafter\let\csname endequation*\endcsname\relax
\usepackage{amsmath}
\usepackage{iopams}  %Uncomment this line if AMS fonts required
\usepackage{graphicx}
\usepackage{subfigure}
\usepackage{float}
\usepackage{color}

 \newcommand{\ket}[1]{|#1\rangle}

\begin{document}

\title[]{Non-reciprocal few-photon devices based on chiral waveguide-emitter couplings}
\author{C~Gonzalez-Ballestero$^{1}$, Esteban~Moreno$^{1}$, F~J~Garc\'ia-Vidal$^{1,2}$, A~Gonzalez-Tudela$^{3}$}
\address{$^{1}$Departamento de F\'isica Te\'orica de la Materia
Condensada and Condensed Matter Physics Center (IFIMAC), Universidad Aut\'onoma de Madrid, E-28049 Madrid,
Spain}
\address{$^2$Donostia International Physics Center (DIPC), E-20018 Donostia/San Sebastian, Spain}
\address{$^{3}$Max-Planck-Institut f\"ur Quantenoptik, Hans-Kopfermann-Strasse 1, 85748 Garching, Germany}

\ead{carlos.ballestero@uam.es}
\ead{alejandro.gonzalez-tudela@mpq.mpg.de}

\begin{abstract}
We demonstrate the possibility of designing efficient, non reciprocal few-photon devices by exploiting the chiral coupling between two waveguide modes and a single quantum emitter. We show how this system can induce non-reciprocal photon transport at the single-photon level and act as an optical diode. Afterwards, we also show how the same system shows a transistor-like behaviour for a two-photon input. The efficiency in both cases is shown to be large for feasible experimental implementations. Our results illustrate the potential of chiral waveguide-emitter couplings for applications in quantum circuitry. 
\end{abstract}

\vspace{2pc}

%\submitto{NJP}
\maketitle

\section{Introduction}

The ability to enhance and tailor the interaction between qubits and photons lies at the heart of quantum circuitry and quantum information protocols \cite{obrien07a}. During the last years, a large experimental effort has been devoted to the study of waveguides as suitable photonic devices for this purpose either by coupling them to solid state \cite{laucht12a,lodahl15a} or atomic emitters \cite{vetsch10a,goban13a}. Indeed, the two ends of a waveguide act as natural ports for introducing and extracting information, making these systems basic elements for complex quantum networks \cite{kimble08a}. Moreover,  the two-dimensional confinement of the guided photons not only allows for a large qubit-field interaction but also facilitates the miniaturization of devices and, more recently, it has allowed the generation of light-matter chiral couplings \cite{mitsch14a,petersen14a,coles15a,sollner15a}, which opens new interesting possibilities in waveguide quantum optics \cite{lodahl16}. On the theoretical side, the interaction between quantum emitters and waveguides has been exploited to design basic 
operations on photonic qubits, e.g., such as single-photon transistor \cite{chang07b} or phase gates \cite{zheng13a,ralph15a}; or as mediators of interactions between qubits for, e.g., entanglement generation  \cite{dzsotjan10a,gonzaleztudela11a,ramos14a,pichler15a,gonzalezballestero15a},  designing quantum gates \cite{paulisch16a} or preparing non-classical states of light \cite{gonzaleztudela15a,gonzaleztudela16a} among others.

Among the wide range of useful optical devices for quantum circuitry, those whose behavior is intrinsically non-reciprocal are especially interesting and challenging to devise as waveguide systems lack of time-reversal symmetry breaking \cite{jalas13a}. The simplest element in this group is the single-photon diode or isolator, in which the propagation of light in different directions is inequivalent or, in an ideal situation, totally suppressed in one of them.  Partially asymmetric transmission has been proposed for systems such as plasmonic waveguides \cite{huang15a}, cavity arrays \cite{weibin15a, mascarenhas14a,shen16a}, or single cavity resonators  \cite{hafezi12a,rosenblum16a}.

With the recent  experimental advances in waveguide fabrication and integration of quantum emitters \cite{laucht12a,lodahl15a,vetsch10a,goban13a}, the design of non-reciprocal photonic elements has experienced a renewed interest with several theoretical proposals either using non-chiral couplings and using two qubits \cite{fratini16a, dai15a,mascarenhas16a,chen15a,fratini14a} or $V$-level systems \cite{yuan15a}; or, exploiting chiral light-matter couplings with a single quantum dot \cite{sollner15a} or atomic ensembles \cite{sayrin15a}, the latest showing experimental isolations of $\sim 8$ dB for $N\approx 30$ atoms.

In this manuscript, we continue along the path of exploiting chiral light-matter couplings for the design of nonreciprocal few-photon circuitry. Our system specifically make use of quantum interference to cancel undesired photonic paths, thus leading to particularly robust and efficient devices by using a single quantum emitter in a $\Lambda$ configuration chirally coupled to two waveguides. In Section \ref{secHAMILT}, we introduce the four-port device under consideration as well as its Hamiltonian. We continue by solving the single-photon scattering for such system in Section \ref{secSINGLE} and showing its behavior as a single-photon rectifier or router, a device which efficiently transfers a photon from one waveguide into another. Additionally, we illustrate how the same setup can be employed as a single-photon diode, which allows a photon to be transmitted only when it travels along a certain direction. After this, we study the scattering of two-photons in this device in Section \ref{sec2PHOTON}, 
demonstrating how a transistor-like behavior is obtained also for realistic parameters.  Finally, our conclusions are presented in Section \ref{secConclusions}.

\section{The model system}\label{secHAMILT}

\begin{figure} 
\center
		\includegraphics[width=\linewidth]{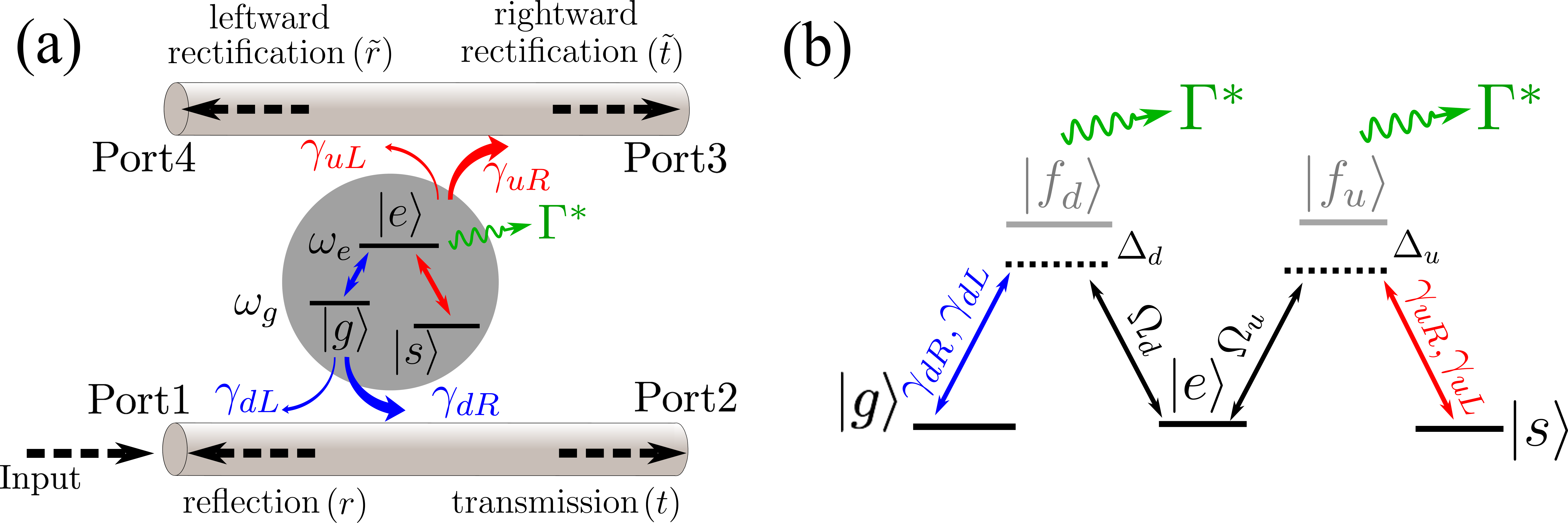}
		%\vspace{-0.2cm}
		\caption{(a) Scheme of the system under study. A three level system in $\Lambda$ configuration interacts with two independent waveguides, labeled $u$ and $d$. The transition $\vert g \rangle \leftrightarrow\vert e\rangle$, depicted in blue, is chirally coupled to the right- and left- propagating photons of the  bottom waveguide, with coupling rates $\gamma_{dR}$ and $\gamma_{dL}$ respectively. The second transition,  $\vert s \rangle \leftrightarrow\vert e\rangle$ (in red) is in turn chirally coupled to the  upper waveguide, with coupling rates $\gamma_{uR}$ and $\gamma_{uL}$. Finally, the excited state $\vert e \rangle$ may decay radiatively into free space modes at a rate $\Gamma^*$. The usual transmission and reflection amplitudes are named $t$ and $r$ respectively, whereas the processes by which the photon is rectified into the second waveguide have scattering coefficients $\tilde{t}$ and $\tilde{r}$, corresponding to right and left propagating photons respectively. (b) Inverted W-system in which two optically excited states $\ket{f_{d,u}}$ are connected to $\ket{e}$ through an off-resonant classical field with amplitude $\Omega_{d,u}$ and detuning $\Delta_{d,u}$, and to $\ket{g}/\ket{s}$ through the lower/upper waveguide. When $|\Delta_{d,u}|\gg\Omega_{d,u}$, the system is equivalent to that of panel (a), 
with renormalized coupling strengths $\gamma_{i\nu}\rightarrow \frac{|\Omega_{i}|^2}{\Delta_i^2}\gamma_{i\nu}$ (and spontaneous emission $\Gamma^*\rightarrow \sum_i\frac{|\Omega_{i}|^2}{\Delta_i^2} \Gamma^*$). }
		\label{fig1}
\end{figure}

The system under study is depicted in Fig. \ref{fig1}a: two waveguides, which we label $u$ and $d$ respectively, form a four port arrangement in which each input/output port is labeled with the numbers $1$ to $4$ as shown in the figure. Each of the waveguides is coupled to one of the two transitions of a central three-level system (3LS) in a lambda configuration. In principle, we allow both these couplings to be chiral, i.e., the coupling rates to left- and right- propagating photons, labelled $\gamma_{jL}$ and $\gamma_{jR}$ ($j=u,d$) respectively, can be different. Additionally, the excited state of the qubit may decay into radiative modes outside of the waveguides at a rate $\Gamma^*$.

%\vspace{-0.6cm}

The Hamiltonian of the system is a generalization of the usual expression in the position basis \cite{fan10a,witthaut10a}, and can be separated into five contributions $(\hbar = 1)$,
\begin{equation}\label{Hinit}
H = H_{3LS} + H_{d}+H_{u} + H_{Id} + H_{Iu}.
\end{equation}
Here, the first term describes the bare $\Lambda$ system,
 \begin{equation}
 	H_{3LS} = (\omega_e -i\Gamma^*/2)\vert e \rangle\langle e \vert + \omega_g\vert g \rangle\langle g \vert,
 \end{equation}
 where the non-hermitian contribution $\Gamma^*$ accounts for the spontaneous emission of the excited state $\vert e \rangle$ into other modes different from the waveguides ones, e.g., free space. The origin of energies is taken at the state $\vert s \rangle$ for convenience. The second and third terms in Eq. (\ref{Hinit}) describe the energy of the photonic modes in the two waveguides, given by
 \begin{eqnarray}
 H_{d} &= -iv_g \!\int dx\! \left(c^\dagger_R(x)\partial_xc_R(x)-c^\dagger_L(x)\partial_xc_L(x)\right),\\
 H_{u} &= -iv_g \!\int dy\! \left(b^\dagger_R(y)\partial_yb_R(y)-b^\dagger_L(y)\partial_yb_L(y)\right).
 \end{eqnarray}
 In both of the waveguides, we assume a linear dispersion relation, where the respective group velocities, $v_g$, are be considered equal in this work for simplicity. The operators $c^\dagger_{R(L)}(x)$ and $b^\dagger_{R(L)}(y)$ are the photonic creation operators in the lower and upper waveguide, respectively. Their corresponding action is to create a right(left)-propagating photon at positions $x$ or $y$. Note that Hamiltonians $ H_{d}$ and $ H_{u}$ are completely equivalent, the only difference being a deliberate change in notation for both operators and position coordinates.
This distinction aims to ease the identification of quantities belonging to each of the two independent waveguides. Finally, it is worth noting that, because of the very general form of the above Hamiltonians, the two photonic reservoirs in our problem do not necessarily represent two physically separated waveguides. Indeed, they could for instance account for two different, uncoupled modes propagating in the same waveguide.
 
The last two terms  in Eq. (\ref{Hinit}) represent the coupling between the two waveguides and the 3LS, which takes place at $x = y =0$ . They are expressed as
 \begin{eqnarray}
   H_{Id}&=\sum_{\alpha=R,L}\int dx \delta(x)V_\alpha c^\dagger_\alpha(x)\vert g \rangle \langle e \vert + H.c. ,\\
    H_{Iu}&=\sum_{\alpha=R,L}\int dy \delta(y)W_\alpha b^\dagger_\alpha(y)\vert f \rangle \langle e \vert + H.c. ,
   \end{eqnarray}
    with $\delta$ representing the Dirac delta distribution. In these expressions, we choose the four coupling constants $\lbrace V_R,V_L,W_R,W_L\rbrace$ to be real numbers for simplicity. They are related to the final decay rates into the waveguides through $\gamma_{d\alpha} = V_\alpha^2/v_g,\gamma_{u\alpha} = W_\alpha^2/v_g$ for $\alpha = R,L$. Note that a key feature of this Hamiltonian is that each transition of the 3LS interacts only with one of the waveguides. Specifically, the transition $\vert g \rangle \leftrightarrow\vert e\rangle$ is coupled to the bottom waveguide, whereas the transition $\vert s \rangle \leftrightarrow\vert e\rangle$ is coupled to the upper waveguide. This coupling structure, essential in the rest of our work, does not isolate one waveguide from another, as they can exchange probability through the excited state $\vert e \rangle$. 
    
Before studying the photon scattering, it is useful to introduce three relevant quantities which will determine the behavior of the system. First, we define the total coupling strength of each transition of the 3LS, $\gamma_j = \gamma_{jR} + \gamma_{jL}$ $(j=d,u)$, which accounts for the total decay rate of the excited state $\vert e\rangle$ into each of the waveguides. The total couplings are used to define the directionalities of each transition, 
\begin{equation}
    D_j = \frac{\gamma_{jR}-\gamma_{jL}}{\gamma_j} \hspace{1.2cm} (j=d,u),
 \end{equation}
which quantify the asymmetry in the 3LS-waveguide couplings. For non-chiral interactions $D_j=0$, whereas for maximally asymmetric coupling $D_j = \pm 1$.  The third relevant magnitude is the Purcell factor, which accounts for the modification of the total decay rate of an emitter when placed in the vicinity of a nanostructure,
\begin{equation}\label{PF}
    P_F = \frac{\gamma_d + \gamma_u}{\Gamma^*_0}.
    \end{equation}
In the equation above, $\Gamma^*_0$ represents the decay rate of the 3LS in vacuum, which we can approximate as $\Gamma^*_0 \approx \Gamma^*$.
The Purcell factor, as well as the related beta factor $\beta = 1-(P_F+1)^{-1}$, are the typical figures of merit in waveguide systems.  

Finally, it is interesting to mention that when several hyperfine and excited levels are available, as it occurs for atomic systems, one can think of an alternative implementation of a $\Lambda$ system that allows for an independent control of the total couplings $\gamma_u$ and $\gamma_d$. One example can be the one depicted in Fig.~\ref{fig1}b where two optically excited states levels $\ket{f_{1,2}}$ are connected to both $\ket{g,s}$ respectively through the lower/upper waveguide. Moreover, the states $\ket{f_{1,2}}$ are also connected with two off-resonant classical lasers with amplitude $\Omega_{1,2}\ll |\Delta_{1,2}|$. Under these conditions, the excited states can be adiabatically eliminated giving rise to an effective dynamics as in Fig. \ref{fig1}a, with renormalized waveguide decay rates $\gamma_{i\nu}\rightarrow \frac{|\Omega_{i}|^2}{\Delta_i^2}\gamma_{i\nu}$ and spontaneous emission $\Gamma^*\rightarrow \sum_i\frac{|\Omega_{i}|^2}{\Delta_i^2} \Gamma^*$. Notice that the directionality parameter $D_j$ is unaltered by this renormalization, whereas 
the Purcell factor only gets a factor half smaller as the spontaneous emission gets also renormalized by the Raman factor. The advantage of this method relies on the couplings to the two waveguides being now fully tunable through $\Omega_{1,2}$, and the states $\ket{g,s,e}$ being long-lived.

%\textcolor{red}{Quiza este ultimo parrafo lo podemos poner en la pag.3, como segundo parrafo de la seccion 2? Asi describimos el sistema, una possible experimental implementation y ya nos metemos con la teoria... Quiza asi quedaria mas integrado en el texto?}

%  \begin{figure} 
%  \center
%  		\includegraphics[scale=0.15]{figures/fig2}
%  		%\vspace{-0.2cm}
%  		\caption{General definition of the scattering coefficients. The usual transmission and reflection amplitudes are named $t$ and $r$ respectively, whereas the processes by which the photon is rectified into the second waveguide have scattering coefficients $\tilde{t}$ and $\tilde{r}$, corresponding to right and left propagating photons respectively. }\label{fig2}
%  \end{figure}  

\section{Single photon devices}\label{secSINGLE}

The complete Hamiltonian of Eq. (\ref{Hinit}) can be fully diagonalized in the single-excitation subspace. In order to study the single-photon scattering, we can restrict the problem to a photon incoming from an arbitrarily selected port, in this case port $1$. The solutions corresponding to an input through ports $2$ to $4$ are not detailed here, as their calculation follows an analogous procedure.

\subsection{Scattering of single photons}

 Our aim is to determine the scattering coefficients for a monochromatic photon incoming through port $1$. Note that if the 3LS is initialy in the state $\vert s \rangle$, it does not interact with the photons in the bottom waveguide, and the scattering solution is reduced to an unperturbed wave travelling from port $1$ to port $2$. Henceforth, our interest is focused on the situation in which the 3LS is initially in the state $\vert g \rangle$. In this situation, the photon can be scattered into four different ports, and we must define four scattering coefficients which are schematically depicted in Fig. \ref{fig1}.
 
The diagonalization of the Hamiltonian in the single-excitation subspace is detailed in Appendix A. The single-photon solution is completely determined by the four scattering coefficients defined in Fig. \ref{fig1}a, which are the probability amplitudes for each of the possible scattering processes. They are given by
 \begin{equation}\label{t}
 t(\omega) = \frac{\omega - \omega_{eg} +  i\Gamma^*/2 +i(\gamma_{dL}-\gamma_{dR}+\gamma_{uL}+\gamma_{uR})/2}{\omega - \omega_{eg} +  i\Gamma^*/2 +i(\gamma_{dL}+\gamma_{dR}+\gamma_{uL}+\gamma_{uR})/2},
 \end{equation}
 \begin{equation}\label{r}
 r(\omega) = \frac{-i\sqrt{\gamma_{dR}\gamma_{dL}}}{\omega - \omega_{eg} +  i\Gamma^*/2 +i(\gamma_{dL}+\gamma_{dR}+\gamma_{uL}+\gamma_{uR})/2},
 \end{equation}
 \begin{equation}\label{ttilde}
 \tilde{t}(\omega) = \frac{-i\sqrt{\gamma_{dR}\gamma_{uR}}}{\omega - \omega_{eg} +  i\Gamma^*/2 +i(\gamma_{dL}+\gamma_{dR}+\gamma_{uL}+\gamma_{uR})/2},
 \end{equation}
 \begin{equation}\label{rtilde}
 \tilde{r}(\omega) = \frac{-i\sqrt{\gamma_{dR}\gamma_{uL}}}{\omega - \omega_{eg} +  i\Gamma^*/2 +i(\gamma_{dL}+\gamma_{dR}+\gamma_{uL}+\gamma_{uR})/2},
 \end{equation}
 where $\omega$ is the energy of the incoming photon, and $\omega_{eg} = \omega_e - \omega_g$ is the energy of the transition $\vert g \rangle \leftrightarrow \vert e \rangle$. It is straightforward to check that the probability is conserved as $\vert t\vert^2 + \vert r\vert^2 + \vert \tilde{t}\vert^2 + \vert \tilde{r}\vert^2 = 1$ when $\Gamma^* = 0$.

\subsection{Single-photon rectifier}

In this section we will show how to tune the system parameters to devise a single-photon router or rectifier, able to direct the input photon from port 1 to port 3 instead of continuing in the same waveguide. For a clearer interpretation of the physical mechanisms involved, let us consider for now the ideal case in which the couplings are maximally chiral and the losses of the 3LS are negligible, i.e., $D_j = 1$ and $P_F \to \infty$ (or equivalently, $\gamma_{dL}=\gamma_{uL}=0$ and $\Gamma^*=0$). In this simple situation, the incoming photon can only be scattered rightwards, and consequently both coefficients $r$ and $\tilde{r}$ vanish. The remaining two scattering amplitudes become
 \begin{equation}
 t(\omega) = \frac{\omega - \omega_{eg} +i(\gamma_{uR}-\gamma_{dR})/2}{\omega - \omega_{eg}  +i(\gamma_{dR}+\gamma_{uR})/2},
 \end{equation}
  \begin{equation}
  	\tilde{t}(\omega) = \frac{-i\sqrt{\gamma_{dR}\gamma_{uR}}}{\omega - \omega_{eg}  +i(\gamma_{dR}+\gamma_{uR})/2}.
  \end{equation}
  From the formulas above, it is straightforward to see that when the frequency of the incoming photon is resonant ($\omega = \omega_{eg}$) and the two remaining couplings are chosen equal ($\gamma_{dR} = \gamma_{uR}$), the transmission coefficient $t$ also vanishes. In this particular situation, three out of the four scattering amplitudes cancel out ($r=\tilde{r}=t=0$), and the incoming photon is directed to Port $3$ with probability $\vert \tilde{t}\vert^2  =1 $. 
  
   \begin{figure} 
   	\includegraphics[width = \linewidth]{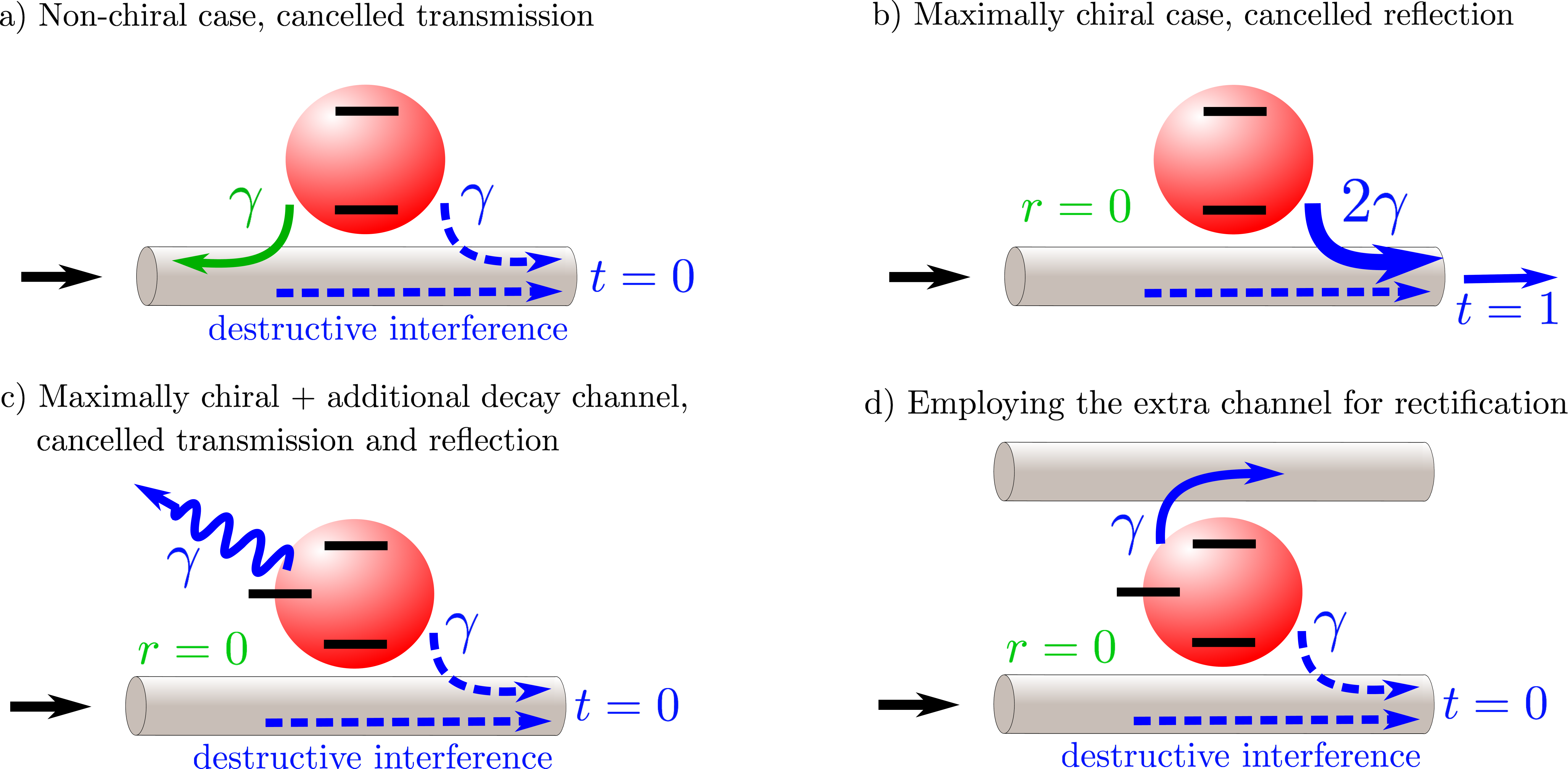}
   	%\vspace{-0.2cm}
   	\caption{a) The single-photon transmittance for a qubit non-chirally coupled to a waveguide vanishes due to a destructive interference. b) When the coupling is maximally chiral, however, the reflection is cancelled and the balance between the previously interfering amplitudes is broken, resulting in full transmission. c) If an extra decay channel is added to the qubit, perfect interference can be achieved again, and both transmission and reflection are canceled. d) Our scheme uses a second waveguide to collect the photon emitted through the extra channel, achieving full rectification.}\label{fig3}
   \end{figure}  
  
  In order to make clearer the underlying physical mechanism of rectification, we first recall the situation of a 2LS symmetrically coupled to a single waveguide, as shown in Fig. \ref{fig3}a, where it is well known that an incoming photon whose frequency is resonant with that of the 2LS is reflected with probability $1$~\cite{fan10a}.  Such perfect reflection is a direct consequence of a destructive interference between the direct transmission and the photon reemitted after absorption The amplitudes of these two processes, shown in dashed lines in Fig. \ref{fig3}a, cancel out as they are equal in magnitude an opposite in sign. 
  
  The situation can be turned around when we allow the qubit-waveguide coupling to be chiral, as Fig. \ref{fig3}b shows. Whenever a photon is absorbed by the 2LS, the chiral interaction introduces an imbalance between the right- and left- reemission probabilities. Hence, while the amplitude of the direct transmission process (dashed blue line) remains unchanged, the absorption+rightward reemission amplitude (solid blue line) increases or decreases in magnitude with respect to the non-chiral situation.  In Fig. \ref{fig3}b, the maximally chiral limit is displayed, where the coupling asymmetry is pushed to its maximum, i.e., no photons can be emitted leftwards.
  Hence, since the reflection of the photon at resonant frequency is impossible, the rightward emission amplitude (thick blue 
line) is now maximized in magnitude, and the transmission probability tends to unity. Chirality thus allows for a complete inversion of the scattering output as compared to the non-chiral case of Fig.~\ref{fig3}a.

Interestingly, it is possible to cancel out \textit{both} transmission and reflection coefficients by adding an extra decay channel (Fig.~\ref{fig3}c). Here, the coupling to left-propagating photons is again set to $0$, but we now allow the excited state to decay into a second and in principle arbitrary environment. If we now choose the two decay rates to be equal as shown in the figure, 
only half of the probability absorbed into the excited state will decay back into rightward guided modes. But as the discussion in Fig.~\ref{fig3}a revealed, this is exactly the fraction of reemitted probability which leads to perfectly destructive interference in transmission. Hence, the transmission coefficient is $0$ again and, having no option of being either reflected or transmitted, the incoming photon is redirected into the secondary environment with maximum probability. The only remaining task in order to recover our four port system is to assume that the extra environment is a second waveguide, as depicted in Fig. \ref{fig3}d. With this addition we introduce the possibility of addressing the rectified photon to a particular port for further use. 

The rectification device is thus achieved by cancelling both transmission and reflection coefficients, therefore forcing the photon to \textit{switch} into the second waveguide. Note, however, that the vanishings of $r$ and $t$ respond to very different causes, in the first case to chirality alone (through $\gamma_{dL}=0$), and in the second to destructive quantum interference. In any case, chirality is essential both to extract the photon from the initial waveguide and to redirect it to the selected output port after the rectification. Similar quantum interference effects, not based on chirality, have been exploited previously in the literature to, e.g.,  enhance photon blockade \cite{bamba11a,majumdar12a,shi13a} or achieving deterministic down-conversion of photon pairs \cite{koshino09a,chang15a,sanchezburillo16a}.

Let us now study the performance of the single-photon rectifier in a more realistic situation, in which the device operation is affected by losses $\Gamma^* \ne 0$ as well as imperfect directionalities $D_j<1$. In principle, we consider the four coupling rates $\gamma_{j\alpha}$ to be different in this case.
First of all, note that even in this general situation we can tune the system parameters so that the transmission coefficient vanishes. Indeed, from Eq. (\ref{t}) it is straightforward to see that $t=0$ for an incoming photon in the resonance condition $(\omega = \omega_{eg})$ whenever the couplings fulfill
\begin{equation}\label{condt0}
\gamma_{dR} = \gamma_{dL}+\gamma_{uL}+\gamma_{uR}+\Gamma^*.
\end{equation}
Note, however, that this condition is limited by physical constraints, and cannot be always achieved. Indeed, if we rewrite Eq. (\ref{condt0}) in terms of Purcell factor and directionalities, 
\begin{equation}
\gamma_u = \gamma_{dR}-\gamma_{dL} - \Gamma^* = \gamma_d\frac{D_d P_F-1}{P_F+1},
\end{equation}
it is clear that a physical solution (i.e. $\gamma_d,\gamma_u>0$) requires the Purcell factor to fulfill
\begin{equation}\label{Purcellcond}
P_F \ge  \frac{1}{D_d}.
\end{equation}
In other words, there is a threshold for the Purcell factor above which the rectification condition $t=0$ can be achieved. The reason behind this fundamental constraint relies on the aforementioned destructive interference, which requires half of the probability emitted in the decay of $\vert e \rangle$ to be directed towards port $2$. If the losses $\Gamma^*$ are so large as to represent more than half of the decay rate of $\vert e \rangle$, there is no possible way of distributing the couplings $\gamma_{jR},\gamma_{jL}$ in order to fulfill this requirement. Equation (\ref{Purcellcond}) thus determines the regime of operation of the single-photon rectifier.
 
 In practical terms, the limitation expressed by Eq. (\ref{Purcellcond}) is not very restrictive for a wide variety of realistic systems. Indeed, for perfectly directional couplings ($D_j = 1$) we can achieve the rectification condition $t=0$ for Purcell factors as low as $1$, whereas for usual experimental values of $0.8 < D_j < 0.95$ \cite{sollner15a} the limit only increases up to $P_F \ge 1.25$. These Purcell factors are very common in most waveguide systems, where values as high as $P_F \sim 30$ have been reported \cite{kolchin15a}. Therefore, from now on we consider the case in which the assumption $t=0$ is fulfilled. By doing so, the only two factors decreasing the performance of the rectifier will be the losses $\Gamma^*$, and the leakage into ports $1$ and $4$ caused by imperfect directionalities. Finally, note that the efficiency of the device can also be diminished if the incoming photon is detuned with respect to the transition frequency $\omega_{eg}$, a situation in which the transmission towards port $2$ 
would not completely vanish. However, this is a minor problem as compared to the finite directionalities and the free-space losses. Indeed, the effect of the detuning is only relevant if such detuning is large as compared to the emission linewidth of the state $\vert e \rangle$, namely $\gamma_d+\gamma_u$. However, for the system to behave as a rectifier, we must tune the coupling rates to fulfill Eq. (\ref{Purcellcond}), a condition that can be also written as $\gamma_d+\gamma_u>\Gamma^*(D_d^{-1}-1)$. Therefore, the emission linewidth of the state $\vert e \rangle$ is always relatively large in the rectifier, making it intrinsically robust against small variations of the resonance condition $\omega = \omega_{eg}$.

 \begin{figure}
 \center
 	\includegraphics[width=\linewidth]{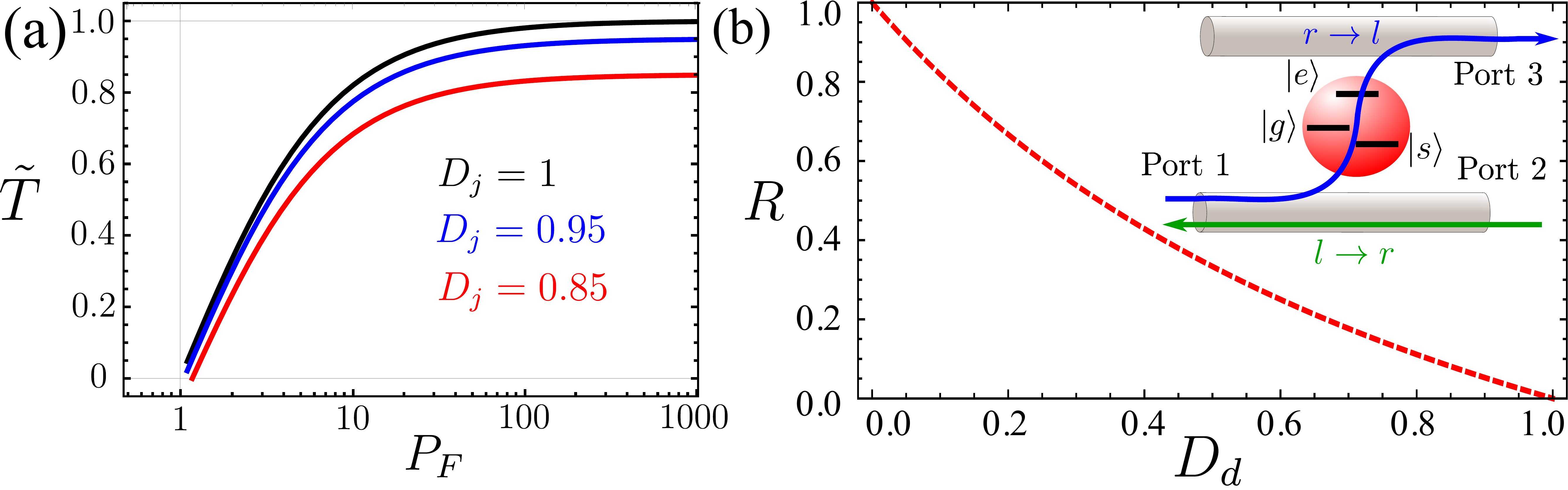}
 	%\begin{picture}(0,0)
 	%\put(-180,38){\includegraphics[scale=0.42]{figures/tvsPURCELLinset2}}
 	%\end{picture}
 	\caption{(a) Total probability of rectification into port $3$ versus Purcell factor, for different directionalities $D_d = D_u$.  (b) Reflection probability determining the efficiency of the diode, versus directionality $D_d$. Inset. Rectifier acting as a single-photon diode with respect to the bottom waveguide. A photon introduced through port $1$ is rectified with probability $1$, and cannot reach port $2$. On the other hand, a photon in port $2$ is transmitted to port $1$ with maximum probability. In both panels we fix $t=0$.}\label{fig4}
 \end{figure}
 
 The efficiency of the single-photon rectifier can be quantified through the total rectification probability which, under the condition $t=0$, is given by
 \begin{equation}
 \tilde{T}=\vert\tilde{t}\vert^2 = \frac{1+D_u}{1+D_d}\frac{D_dP_F-1}{P_F+1} \;\;\;\; \;\;\;\;\;\;\;\; \left(\text{for $P_F\ge \frac{1}{D_d}$}\right)
 \end{equation} 
 for a photon in the resonance condition. Note that in the ideal case ($P_F \to \infty$ and $D_j \to 1$) the efficiency defined above is equal to $1$, whereas in a realistic case the probability leakage into the undesired channels (free-space, as well as ports $1$ and $4$) will reduce this value. %Interestingly, this quantity does not depend on the individual coupling strength, $\gamma_j$, thus making the following results applicable to a wide range of waveguide systems. 
 The scattering probability $\vert \tilde{t}\vert^2$ is displayed in Fig. \ref{fig4}a as a function of the Purcell factor and for different values of the directionalities $D_d$ and $D_u$, considered equal for simplicity. The rectification probability  is shown to remain rather close to unity for realistic directionalities, for instance as high as $\sim80\%$  for easily achievable values of $P_F = 15$, $D_j = 0.9$. 

For completeness, let us mention that it is possible to relax the requirements for non-reciprocal transport if the only important thing is to block one of the direction of propagation, i.e., the optical isolator or \emph{diode} configuration as defined in reference \cite{sayrin15a}. For example, let us study the efficiency of the system to act as   a single photon diode with respect to the bottom waveguide, as depicted in Fig. \ref{fig4}b. Here, we can define two different paths for the single photon, namely a photon incoming from port $1$ towards port $2$, and the opposite situation in which the photon is introduced through port $2$ towards port $1$. We will name these paths $l\to r$ and $r \to l$ respectively. Due to the chiral coupling, the single-photon scattering coefficients are \textit{different} for these two paths. 
On the one hand, for the path $l\to r$ the scattering amplitudes have already been calculated in Eqs. \ref{t}-\ref{rtilde}. Here, if the rectification condition $t=0$ is fulfilled, a photon from port $1$ can never reach port $2$ since it is rectified into the second waveguide. On the other hand, for the path $r\to l$ the scattering coefficients can be calculated in the same fashion and, for a photon in the resonance condition $\omega = \omega_{eg}$, they are equal to
$
t_{r\to l} = 1-R_{l\to r}
$, 
$
 r_{r\to l} = r_{l\to r}
$,
$
\tilde{t}_{r\to l} = \tilde{t}_{l\to r}\sqrt{\gamma_{dL}/\gamma_{dR}}
$, and $
\tilde{r}_{r\to l} = \tilde{r}_{l\to r}\sqrt{\gamma_{dL}/\gamma_{dR}}
$ respectively.
 With the exception of photon transmission to port $1$, all the processes in these expressions explicitly require the absorption of the left-propagating photon by the 3LS, and are therefore proportional to $\gamma_{dL}$. In the ideal case of perfect directionalities we have $\gamma_{jL}=0$, and thus the photon is always transmitted to port $1$. As a consequence, in the ideal situation and under the rectification condition the system fulfills
\begin{equation}
t_{l\to r} = 0 \hspace{1cm} ; \hspace{1cm} t_{r\to l} = 1,
\end{equation}
which is by definition the behavior of a single-photon diode.

In a realistic case, the performance as a diode is even better than as a rectifier, since its operation imposes less restrictive conditions on the route covered by the incoming photon, as we will see below
. Let us study the operation along the two different paths, assuming that the $t=0$ condition, Eq. (\ref{condt0}), is fulfilled. First, when the photon is sent along the path $l\to r$, it cannot be transmitted to port $2$ since $t=0$, and therefore still perfectly fullfills the desired behavior for a diode.  The only decrease in performance in this situation originates from reflections back into port $1$, which can introduce noise in the device. Hence, the efficiency of the diode along this photonic path is determined exclusively by the reflection probability
 \begin{equation}\label{R}
      R_{l\to r} \equiv R = \frac{1-D_d}{1+D_d} \;\;\;\;\; \left(\text{for } D_d >0 \text{ and } P_F\ge \frac{1}{D_d}\right).
      \end{equation}   
 On the other hand, for a photon incoming along the opposite path $r\to l$, two sources of loss arise, namely  a possible reflection back into port $2$, and photon leakage into either free space or the upper waveguide. The overall effect of such losses is to reduce the total transmission probability below $1$. Hence, the total efficiency along the path $r\to l$ is determined by the transmission probability $T_{r\to l}$ which, after manipulation, can be shown to be
 \begin{equation}\label{Tdiode}
T_{r\to l} = \left(1-R_{l\to r}\right)^2= \left(1-R\right)^2 \;\;\;\;\;\;\;\;\;\; \left(\text{for $P_F\ge \frac{1}{D_d}$}\right).
 \end{equation}    
 According to equations (\ref{R}) and (\ref{Tdiode}), the performance of the diode is a function \textit{only} of the directionality $D_d$, through the reflection probability $R_{l\to r} \equiv R$. This is easily understood for the path $r \to l$, where all the undesired processes depend on $\gamma_{uL}$ (and thus on $D_d$) as we have seen above. On the other hand, for the path $l\to r$ we only need the photon to be extracted from the waveguide, but its final destination (namely free space modes, port $3$, or port $4$) is irrelevant. The operation of the diode is thus not dependent on the the particular value of $\Gamma^*$, $\gamma_{uR}$, and $\gamma_{uL}$, but on the total external loss rate $\gamma_{uR}+\gamma_{uL} + \Gamma^*$. Since such rate is related to $D_d$ through the $t=0$ condition (see Eq. \ref{condt0}), the performance of the diode depends exclusively on the parameter $D_d$. The reason behind the diode being more robust relies on the less restrictive conditions for its operation, specifically regarding the route of the photon incoming through port $1$.
 Whereas for the diode it is enough to extract such photon from the waveguide $d$, the rectifier additionally requires it to be addressed to a given port of waveguide $u$. For this reason, any deviation from the ideal conditions will affect the rectifier in a more drastic way. In Fig. \ref{fig4}b we characterize the losses of the diode by displaying the probability $R$ versus the directionality $D_d$. For directionalities $D_d\gtrsim 0.9$ the reflection losses are very low, $R \sim 5\%$, and the transmission probability along the path $r\to l$ remains at $T_{r\to l} \sim 90\%$.

\section{Two-photon transistor}\label{sec2PHOTON}

In this section we will first characterize the two-photon response of our system by calculating the two-photon wavefunction, and study how it also shows non-reciprocal features. In particular, we will study how this device shows a transistor-like behaviour  \cite{chang07b} when two photons arrive simultaneously through port 1, whereas both of them are transmitted when impinging through port 2. 

\subsection{Scattering of a two-photon state.}

There exist several methods to calculate the multiphoton response of non-linear systems such as LSZ reduction \cite{shi09a} or input-output formalism \cite{shi15a,caneva15a,xu15a}. In this manuscript, we choose to diagonalize directly the Hamiltonian Eq.~(\ref{Hinit}) in the two-excitation subspace. We have checked the consistency of the results with $S$-matrix calculations using input/output methods \cite{shi15a}.

       %\begin{figure}
       	%\includegraphics[width=\linewidth]{fig5.pdf}
       	%\caption{Schematic idea for the single-photon transistor. a) Definition of the usual ports in the notation of a FET transistor, plus an extra auxiliary port. b) Provided that the rectification condition is fulfilled, a single photon from the Source terminal can never reach the Drain. c) A Gate photon makes use of the rectification to change the state of the 3LS from $\vert f \rangle$ to $\vert g \rangle$, thus allowing for the Gate photon to be addressed to the Drain terminal.}\label{fig5}
       %\end{figure}       
First of all, we need to solve the scattering eigenstate associated with a two-photon input, i.e., two waves with well defined momentum $k_{1}$ and $k_2$ incoming through Port $1$, the initial state of the 3LS being $\vert g \rangle$. Following the same steps as in the single excitation subspace, we define the general two-excitation eigenstate for our problem,
 \begin{equation}\label{2photANSATZ}
 \begin{split}
 \vert \epsilon\rangle & = \int dx_1\int dx_2\; \left( \sum_{\alpha=R,L}\phi_{\alpha\alpha}(x_1,x_2) c^\dagger_\alpha (x_1)c^\dagger_\alpha(x_2)  + \phi_{RL}(x_1,x_2) c^\dagger_R (x_1)c^\dagger_L(x_2)\right)\vert g \rangle\\
 +&   \int dx\int dy\; \sum_{\alpha,\beta}\psi_{\alpha\beta}(x,y) c^\dagger_\alpha(x) b^\dagger_\beta(y)\vert s \rangle +  \int dx \sum_{\alpha=R,L} \varphi_\alpha(x)c^\dagger_\alpha(x)\vert e\rangle.
 \end{split}
 \end{equation}
 In the above equation, the wavefunctions $\phi_{\alpha\beta}$ correspond to states in which both photons are in the bottom waveguide. Two of these functions are subject to a bosonic symmetry constraint $\phi_{\alpha\alpha}(x_1,x_2) = \phi_{\alpha\alpha}(x_2,x_1)$. The wavefunctions $\psi_{\alpha\beta}$ describe states with one photon in each of the waveguides, whereas the functions $\varphi_\alpha$ account for states in which one of the excitations is in the state $\vert e \rangle$ of the 3LS. The explicit calculation of the wavefunctions above is detailed in \ref{appendixB}. 
 
 The two-photon wavefunctions have a complicated form, their scattering outputs being thus not straightforward to quantify. Instead of particular scattering coefficients, we will make use of the general detection probabilities $P_{mn}$, which represent the total probability of detecting one photon in port $m$ and another photon in port $n$ after the scattering event occurs. In order to calculate these quantities, we will follow a similar procedure as in Ref. \cite{zheng10a}. We start by splitting the above eigenstate into two contributions,
 \begin{equation}
 \vert \epsilon \rangle = \vert \epsilon_i\rangle+ \vert \epsilon_o\rangle.
 \end{equation}
 The first term in the above equation is the input state $\vert \epsilon_i\rangle$, formed by all the terms in Eq. \ref{2photANSATZ} containing a right-propagating photon in $x<0$ (see \ref{appendixC} for details). The remaining contributions form the scattering output state $\vert \epsilon_o \rangle$.

Let us briefly summarize the definition of the detection probabilities $P_{mn}$ by using a particular example, namely $P_{23}$, and leave the general calculation of all the $P_{mn}$ to \ref{appendixC}. The photons detected at port $2$ will be those propagating rightwards in the bottom waveguide. In the same fashion, photons addressed to port $3$ are right-propagating modes of the upper waveguide. Therefore, we can write the position probability density associated with one photon in port $2$ and another in port $3$ as the following second order correlation function
\begin{equation}
\rho_{23}(x,y) = \frac{\langle \epsilon_o \vert b_R^\dagger(y)c_R^\dagger(x) c_R(x)b_R(y)\vert\epsilon_o\rangle}{\langle\epsilon_o\vert\epsilon_o\rangle\vert_{\Gamma^*=0}},
\end{equation}
where the normalization constant is fixed to the lossless output state. This normalization is also implicitly used in all the scattering problems solved with this formalism in the literature \cite{zheng10a}. The total probability of detecting two photons in such ports is then straightforward,
\begin{equation}
P_{23} = \int_{-L/2}^{L/2} dx \int_{-L/2}^{L/2} dy \rho_{23}(x,y),
\end{equation}
 %Once the scattering output is defined, we can write the detection probabilities as
 %\begin{equation}\label{Pij}
 %P_{ij} = \frac{1}{1+\delta_{ij}}\frac{1}{\langle \epsilon_o \vert \epsilon_o \rangle}\int_{\delta_i} dz_i \int_{\delta_j} dz_j \Big\vert  a_i (z_i) a_j(z_j) \vert \epsilon_o\rangle\Big\vert^2.
 %\end{equation}
 %In the above expression, $z_m = \lbrace x \text{ for } m = 1,2 \text{ ; } y \text{ for } m = 3,4\rbrace$ is a generic positional coordinate. The operators $a$ and integration domains $\delta$ depend on the indices $i,j$ through
 %\begin{equation}
 %\lbrace a_m , \delta_m\rbrace= 
 %\left\{\begin{array}{lr}
 %c_L, (-L/2,0] &\;\;\;\; \text{     for    } m=1 \\
%c _R, [0,L/2) &\;\;\;\; \text{     for    } m=2 \\
%b_R, [0,L/2) &\;\;\;\; \text{     for    } m=3 \\
%b_L, (-L/2,0] &\;\;\;\; \text{     for    } m=4 \\
 %\end{array}\right.
 %\end{equation}
 where $L$ is the total length of the waveguides, for which the limit $L\to \infty$ is considered in this work. The explicit expressions for all the $P_{mn}$ are calculated in Appendix C.
  
The detection probabilities defined above account for the scattering outputs in two particular ports, $m$ and $n$. Note that, nevertheless, there are additional possible processes in the output state $\vert \epsilon_o\rangle$ which should be taken into account. In particular, the contributions from states in which one of the excitations is in the state $\vert e \rangle$ while the second is a propagating photon, described by the wavefunctions $\varphi_\alpha(x)$. These processes can be relevant for incoming wavepackets whose frequency width is comparable to the intrinsic linewidth of the transition $\vert g \rangle \leftrightarrow\vert e\rangle$ \cite{zheng10a}. However, for monochromatic inputs and in the long waveguide limit we are working on, it is possible to demonstrate that the detection probability for any of these processes is infinitely small as compared to the two-photon probabilities $P_{mn}$ (see \ref{appendixC} for details).  Hence, the processes described by the wavefunctions $\varphi_\alpha(x)$ can be safely ignored in the study of the scattering output.

As for the different scattering outputs, note that although there are $16$ possible combinations of indices ${m,n} \in [1,4]$, not all of them represent independent processes. Indeed, we can reduce the number to $10$ by noticing that some probabilities represent the same scattering output $(P_{mn} = P_{nm})$. Additionally, from the general form of the eigenstate in Eq. (\ref{2photANSATZ}) we can immediately deduce that $P_{33} = P_{34} = P_{44}=0$, as the Hamiltonian does not allow for two photons to be rectified. Therefore, only $7$ possible scattering outputs remain, namely $P_{1n}$ and $P_{2n}$, in which one photon is addressed to port $n$ and the second is reflected or transmitted, respectively. Note, finally, that these are the only possible output processes in the absence of free-space losses $\Gamma^*$, and therefore add up to unity,
  \begin{equation}
  \sum_{m=1,2} \sum_{m=n}^{4} P_{mn} \Big\vert_{\Gamma^*=0}= 1.
  \end{equation}
Under this convention, the probabilities $P_{mn}$ play a similar role in the two-photon scattering process as the scattering probabilities $T,R,\tilde{T},\tilde{R}$ did in the single-photon case.

  \subsection{Operation and performance of the two-photon transistor.}
  
  Let us consider first the ideal case in which the 3LS is lossless and the directionalities are maximized, i.e. $\Gamma^* = 0$ and $D_j=1$. Additionally, we will always assume the frequency of both incoming photons to be resonant with the transition $\vert g \rangle \leftrightarrow \vert e \rangle$. Under these conditions the detection probabilities have very simple expressions,
  \begin{align}
    P_{11} &\propto R^2 = 0 , \\
    P_{12} &\propto RT = 0 , \\
    P_{13} &\propto R\tilde{T} = 0 , \\
    P_{14} &\propto R\tilde{R} = 0 , \\
     P_{24} &\propto \tilde{R}(1+T) = 0,
    \end{align}
  \begin{equation}
  P_{22} = T^2 \;\;\;\;\; , \;\;\;\;\; P_{23} = 1-P_{22}.
  \end{equation}
   Here, we have defined $\lbrace T,R, \tilde{T},\tilde{R}\rbrace \equiv \lbrace \vert t(\omega_{eg})\vert^2,\vert r(\omega_{eg})\vert^2,\vert \tilde{t}(\omega_{eg})\vert^2,\vert \tilde{r}(\omega_{eg})\vert^2\rbrace$, where $t,r,\tilde{t},$ and $\tilde{r}$ are the single-photon scattering coefficients defined in Eqs. (\ref{t}-\ref{rtilde}). Naturally, all the processes involving the reflection coefficients $r$ or $\tilde{r}$ vanish in the limit $D_j =1$, and only two processes remain. First, direct transmission of two photons towards port $2$, with probability $P_{22} = T^2$, and second,
    the process by which one of the photons is rectified into port $3$ and the second is transmitted to port $2$, with probability $P_{23} \propto 1-T^2$.

   The ideal situation described above can be extremely useful
   under the rectification condition discussed in the single-photon case, where the transmission probability $T$ also vanishes if the couplings are adequately tuned. When this condition ($T=0$) is fulfilled, the probability $P_{22}$ also vanishes, and only one possible scattering output remains, namely the one described by $P_{23}$. In other words, there is only one possible path for the two-photon wavepacket, with probability $100\%$. This surprising result can be used to build a transistor-like device. For the sake of comparison with an ordinary three-terminal transistor, let us name port $1$ the Source/Gate and port $2$ the Drain. For a single photon input in port $1$ (the source), transmission towards port $2$ is prevented by the rectification process as discussed in section \ref{secSINGLE}. However, if we introduce a second photon through port $1$, one of the input photons is rectified while the second is addressed to port $2$. In this way, a 
transmission channel between ports $1$ and $2$ can be opened by means of a second Gate photon. A single-photon transistor has therefore been achieved which, in the ideal situation we are considering, has an efficiency of $100\%$. Note that the two-photon response of this device is still non-reciprocal, since a two-photon input introduced through port $2$ would travel unperturbed towards port $1$.

  We now consider the more realistic system in which the directionalities are not perfect and the Purcell factor is finite, i.e., $D_j<1$ and $\Gamma^*\ne 0$. As we have discussed above, the rectification process is a key requirement for the operation of the device. Hence, we will assume again that the system parameters have been tuned to fulfill the rectification condition $T=0$, as described in Section $3$. In this situation, the probabilities $P_{mn}$ can be expressed in terms of both the directionalities and the Purcell factor as
  \begin{equation}\label{Pijcomplex1}
  P_{23} = \frac{1+D_u}{1+D_d}\frac{P_FD_d-1}{P_F+1},
  \end{equation}
  \begin{equation}
  P_{11} = Q_d^2,
  \end{equation}
  \begin{equation}
  P_{13} = Q_d P_{23}  \;\;\;\;\;\; ; \;\;\;\;\;\; P_{24} = Q_u(1+Q_d) P_{23}  \;\;\;\;\;\; ; \;\;\;\;\;\; P_{14} = Q_d Q_u P_{23}  ,
  \end{equation}
  \begin{equation}\label{Pijcomplex2}
  P_{12} = P_{22} = 0,
  \end{equation}
  where we have defined $Q_j = (1-D_j)(1+D_j)$.

       \begin{figure}
       	\includegraphics[width=\linewidth]{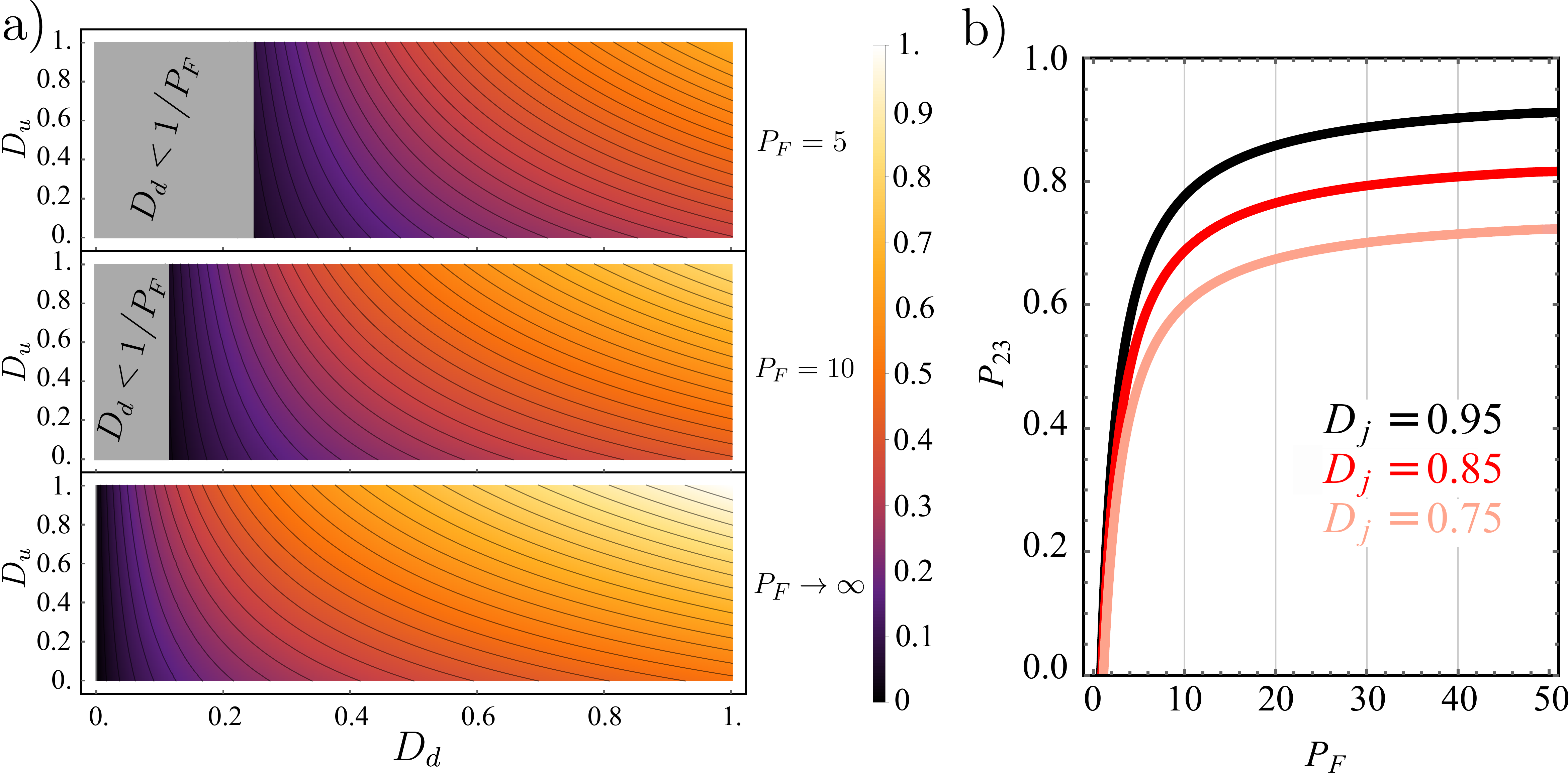}
       	\caption{Performance of the single-photon transistor. a) Success probability $P_{23}$ as a function of the directionalities $D_d$ and $D_u$. Each subpanel corresponds to a different value for the Purcell factor, and its domain is constrained by the fundamental limit Eq. (\ref{Purcellcond}). b) $P_{23}$ versus Purcell factor $P_F$, for different values of the directionalities $D_d = D_u$. In both panels the couplings are tuned to fulfil $t=0$.}\label{fig6}
       \end{figure}
  
  The efficiency of the single-photon transistor is determined by the probability $P_{23}$, which is displayed in Fig.\ref{fig6}a as a function of the two directionalities $D_d\text{ and } D_u$ and for different values of the Purcell factor. Note that, whereas $D_u$ can have any value between $0$ and $1$, the allowed interval of directionalities $D_d$ is restricted by the constraint $T=0$, as described by the condition Eq. (\ref{Purcellcond}). Moreover, the probability $P_{23}$ depends more dramatically on the directionality $D_d$ than on $D_u$. The reason behind this imbalance is that a value $D_u<1$ introduces losses only in the path of the rectified photon, not in the photon traveling towards port $2$. On the other hand, when $D_d$ decreases below $1$, both the transmission and rectification probabilities are affected, increasing the losses of the device in a more drastic way. Therefore, an adequate optimization of $D_d$ is a crucial step towards an efficient photon transistor. In Fig. \ref{fig6}b we 
show the total efficiency as a function of the Purcell factor, for different values of the directionalities $D_d=D_u$. 
The performance of the transistor rapidly approaches $0$ when $P_F$ decreases toward its fundamental limit Eq. (\ref{Purcellcond}). However, such performance saturates to a constant value above a certain Purcell Factor. For experimentally reported values, such as $D_j = 0.9$ and $P_F = 20$, the efficiency of the device reaches  $P_{23} \sim 80\%$. This makes our proposal a feasible device for state of the art experimental setups.

To conclude this section, our most relevant result is the demonstration of chirally coupled systems as highly promising platforms for devising photonic devices beyond the single-photon level. This has been shown by explicitly diagonalizing the system Hamiltonian in the two excitation subspace. Although we have focused on one particular application, namely the transistor, our four-port arrangement is very flexible and could therefore be tuned to perform a wide variety of other operations on the incoming two-photon inputs such as, for instance $50/50$ beam splitters.

 \section{Conclusions}\label{secConclusions}
 
 A new family of few-photon non-reciprocal devices has been presented,
 whose operation is based on quantum interference tuned by chiral waveguide-emitter coupling.
   By studying a simple four-port system, we have shown how an adequate tuning of the parameters can lead to perfect single-photon rectification, an effect we have employed for the design of a single-photon diode.  After, we have analyzed the performance of our device for a two-photon input, demonstrating how a transistor-like behaviour can be achieved. All these devices are shown to operate with high efficiency for experimentally reported parameters. The set of devices that we have introduced represents an additional application of chiral photon-emitter interaction for quantum applications, and provide a flexible and efficient resource for the design and miniaturization of quantum circuits.

 \section*{Acknowledgments}
 A.G.-T. is indebted to T. Shi and Y. Chang for very useful discussions.  CGB acknowledges the Spanish MECD (FPU13/01225 fellowship).  A.G.-T. acknowledges support from the Intra-European Marie-Curie Fellowship NanoQuIS (625955). CGB and FJGV acknowledge the European Research Council (ERC-2011-AdG Proposal No. 290981). EM, FJGV and CGB acknowledge the Spanish MINECO (MAT2014-53432-C5-5-R grant).

 \appendix
 
 \section{Diagonalization in the single-excitation subspace.}\label{appendixA}
 
  Several works contain detailed information on the single-excitation diagonalization of these kind of Hamiltonians \cite{fan10a,gonzalezballestero13a}, hence we will only briefly summarize the key steps. First, we define the general form for our single-excitation eigenstate,
  \begin{equation}\label{Ansatzsingle}
  \begin{split}
  \vert \epsilon \rangle  &= \alpha \vert e \rangle +\int dx\left(\phi_R(x)c^\dagger_R(x)+\phi_L(x)c^\dagger_L(x)\right)\vert g\rangle +\\+ &\int dy\left(\psi_R(y)b^\dagger_R(y)+\psi_L(y)b^\dagger_L(y)\right)\vert s\rangle,
  \end{split}
  \end{equation}
  where the coefficients $\alpha, \phi_\alpha(x),\psi_\beta(y)$ are unknown functions to determine. In order to do so, we solve the time-independent Schr\"odinger equation $H\vert\epsilon\rangle = \epsilon\vert\epsilon\rangle$ by directly applying the Hamiltonian $H$ in Eq. (\ref{Hinit}) to the eigenstate above. In this way, we obtain the following system of equations,
  \begin{equation}
  (\epsilon-\omega_e+i\Gamma^*/2)\alpha = \sum_{\beta=R,L}V_\beta\phi_\beta(0) + W_\beta\psi_\beta(0),
  \end{equation}
  \begin{equation}\label{eqs11}
  (\epsilon - \omega_g + iv_g\partial_x)\phi_R(x) = \alpha V_R \delta(x),
  \end{equation}
  \begin{equation}
  (\epsilon - \omega_g - iv_g\partial_x)\phi_L(x) = \alpha V_L \delta(x),
  \end{equation}
  \begin{equation}
  (\epsilon  + iv_g\partial_y)\psi_R(y) = \alpha W_R \delta(y),
  \end{equation}
  \begin{equation}\label{eqs1last}
  (\epsilon - iv_g\partial_y)\psi_L(y) = \alpha W_L \delta(y),
  \end{equation}
 where we use the short-hand notation $\partial_x \equiv \partial/\partial x$. We now proceed to make an Ansatz for the photonic wavefunctions in terms of scattering coefficients,
  \begin{equation}
  \phi_R(x) = e^{i(\epsilon-\omega_g) x/v_g} \left(\theta(-x) + t\theta(x)\right),
  \end{equation} 
  \begin{equation}
  \phi_L(x) = e^{-i(\epsilon-\omega_g) x/v_g} r\theta(-x) ,
  \end{equation} 
  \begin{equation}
  \psi_R(y) = e^{i\epsilon y/v_g}  \tilde{t}\theta(y),
  \end{equation} 
  \begin{equation}
  \psi_L(y) = e^{-i\epsilon y/v_g} \tilde{r}\theta(-y), 
  \end{equation} 
  which allows for an integration of the system of equations  (\ref{eqs11}-\ref{eqs1last}) around $x = y = 0$ in order to get rid of the delta functions. After such integration, the problem is reduced to a $5\times 5$ system of algebraic equations. The solutions to this system, after the trivial substitution $\omega = \epsilon-\omega_g$, are the scattering coefficients Eqs. (\ref{t},\ref{r},\ref{ttilde},\ref{rtilde}) in the main text.

 \section{Diagonalization in the two-excitation subspace.}\label{appendixB}
 
 The basic steps for the diagonalization in this case are the same as in the single-excitation problem, starting by the general form of the two-excitation eigenstate Eq.(\ref{2photANSATZ}). By directly applying the Hamiltonian (\ref{Hinit}) to the eigenstate $\vert \epsilon\rangle$ we
 obtain a system of differential equations relating all the coefficients. For the sake of compactness, let us first define the variable 
 \begin{equation}
 \eta_\alpha = \Big\lbrace\begin{array}{lcr}
 1 & \text{for} & \alpha = R\\
 -1 & \text{for} & \alpha = L
 \end{array},
 \end{equation}
 as well as the function
 \begin{equation}
 G_{\alpha}(x_1,x_2) = \varphi_\alpha(x_1)\delta(x_2).
 \end{equation}
With these useful definitions at hand, we can express the system of equations in the following form,
\begin{equation}\label{system2first}
\left[\epsilon -\omega_g + iv_g\left(\eta_\alpha\partial_1+\eta_\beta\partial_2\right)\right]\phi_{\alpha\beta} 
=\big(V_\alpha G_\beta(x_2,x_1) + V_\beta G_\alpha(x_1,x_2)\big)\frac{2-\delta_{\alpha\beta}}{2},
\end{equation}
\begin{equation}\label{system2last}
\left[\epsilon  + iv_g\left(\eta_\alpha\partial_x+\eta_\beta\partial_y\right)\right]\psi_{\alpha\beta} = W_\beta G_\alpha(x,y),
\end{equation}
	\begin{equation}
	\left[\epsilon - \omega_e +iv_g\partial_x\right] \varphi_R(x) = 2V_R\phi_{RR}(x,0) + V_L\phi_{RL}(x,0) + W_R\psi_{RR}(x,0) + W_L\psi_{RL}(x,0),
	\end{equation}
	\begin{equation}
	\left[\epsilon - \omega_e -iv_g\partial_x\right] \varphi_L(x) = 2V_L\phi_{LL}(x,0) + V_R\phi_{RL}(0,x) + W_L\psi_{LL}(x,0) + W_R\psi_{LR}(x,0),
	\end{equation}
where $\partial_j \equiv \partial/\partial x_j$ and $\delta_{\alpha\beta}$ represents the Kronecker delta. In a general case the losses are included as an imaginary part in $\omega_e$, i.e. $\omega_e \to \omega_e -i\Gamma^*/2$.

The homogeneous solution to the above differential equations is a two-variable plane wave. The only difficulty is posed by the delta functions, which account for the matching conditions for these waves at the position of the 3LS, $x = y = 0$. It is then necessary to carefully define the different domains in which the functions $\psi,\phi$ are well defined,
\begin{enumerate}
	\item[-] Region (i): $x_1,x_2 <0$, or $x,y,<0$.
	\item[-] Region (ii): $x_1<0<x_2 $, or $x<0<y$.
	\item[-] Region (iii): $0<x_1,x_2$, or $0<x<y$.
	\item[-] Region (iv): $x_2<0<x_1$, or $y<0<x$.
\end{enumerate}
We can do the same for the one-variable functions $\varphi_\alpha(x)$,
\begin{equation}
\varphi_\alpha(x) = \varphi_\alpha^<(x)\theta(-x) + \varphi_\alpha^>(x)\theta(x) .
\end{equation}
Once the different regions are defined, it is possible to simplify the problem by imposing physical restrictions. In particular, as we are interested in the scattering of two photons incoming through port $1$, we can impose the condition that no photons are introduced through other ports. This restriction applies as a series of constraints in our wavefunctions, in particular
\begin{equation}
\begin{split}
\phi_{LL}^{(iii)} & = \phi_{LL}^{(ii)} = \phi_{RL}^{(iii)} = \phi_{RL}^{(ii)} =0, \\
\psi_{RR}^{(i)} & =\psi_{RR}^{(iv)}=0, \\
\psi_{RL}^{(ii)} & =\psi_{RL}^{(iii)}=0, \\
\psi_{LR}^{(i)} & =\psi_{LR}^{(iii)}=\psi_{LR}^{(iv)}=0, \\
\psi_{LL}^{(ii)} & =\psi_{LL}^{(iii)}=\psi_{LL}^{(iv)}=0, \\
\varphi_{L\;}^>  & = 0.
\end{split}
\end{equation}
Finally, we can integrate  Eqs. (\ref{system2first}-\ref{system2last}) around $x,y=0$ to get rid of the delta functions, obtaining the following system of equations and boundary conditions:
\begin{eqnarray}
\big[\omega + iv_g & \left(\eta_\alpha\partial_1+\eta_\beta\partial_2\right)\big]\phi_{\alpha\beta}^{(j)}  =0, \\
& iv_g  \left(\phi^{(ii)}_{RR}(x,0)-\phi^{(i)}_{RR}(x,0)\right)=\frac{V_R}{2}\varphi^<_R(x), \\
 & iv_g \left(\phi^{(iii)}_{RR}(0,x)-\phi^{(ii)}_{RR}(0,x)\right)=\frac{V_R}{2}\varphi^>_R(x)\\
& iv_g  \phi_{LL}^{(i)}(x,0)=\frac{V_L}{2}\varphi_L^<(x),\\
&  iv_g  \phi_{RL}^{(i)}(x,0) = V_L\varphi^<_R(x), \\
&  iv_g  \phi_{RL}^{(iv)}(x,0)= V_L\varphi^>_R(x), \\
&  iv_g  \left(\phi_{RL}^{(iv)}(0,x)-\phi_{RL}^{(i)}(0,x)\right) = V_R\varphi^<_L(x). 
\end{eqnarray}
\begin{eqnarray}
\big[\omega +\omega_g + iv_g & \left(\eta_\alpha\partial_x+\eta_\beta\partial_y\right)\big]\psi_{\alpha\beta}^{(j)}=0, \\
& iv_g\left(\begin{array}{l}
\psi_{RR}^{(ii)}(x,0) \\
\psi_{RR}^{(iii)}(x,0) 
\end{array}\right) = W_R\left(\begin{array}{l}
\varphi_R^<(x) \\
\varphi_R^>(x)
\end{array}\right),\\
& \psi_{RR}^{(iii)}(0,y) =\psi_{RR}^{(ii)}(0,y), \\ 
& iv_g\left(\begin{array}{l}
\psi_{RL}^{(i)}(x,0) \\
\psi_{RL}^{(iv)}(x,0) 
\end{array}\right) = W_L\left(\begin{array}{l}
\varphi_R^<(x) \\
\varphi_R^>(x)
\end{array}\right),\\
& \psi_{RL}^{(iv)}(0,y) =\psi_{RL}^{(i)}(0,y), \\ 
& iv_g\left(\begin{array}{l}
\psi_{LL}^{(i)}(x,0) \\
\psi_{LR}^{(ii)}(x,0) 
\end{array}\right) = \left(\begin{array}{l}
W_L\varphi_L^<(x) \\
W_R\varphi_L^<(x)
\end{array}\right),\\
& \psi_{LR}^{(ii)}(0,y)=\psi_{LL}^{(i)}(0,y)=0.
\end{eqnarray}
\begin{equation}
\begin{split}
\left[\omega - \omega_{eg} +iv_g\partial_x\right] \varphi^>_R(x) = \; \; \; \; \; \; \; \; \; \; \; \; \; \; \; \; \; \; \; \;  \; \; \; \; \; \; \; \; \; \; \; \; \; \; \; \; \; &\\  =V_R\left(\phi^{(iii)}_{RR}(0,x)+\phi^{(ii)}_{RR}(0,x)\right) + V_L\phi^{(iv)}_{RL}(x,0)/2 &+ W_R\psi^{(iii)}_{RR}(x,0)/2 + W_L\psi^{(iv)}_{RL}(x,0)/2.
\end{split}
\end{equation}
\begin{equation}
\begin{split}
\left[\omega - \omega_{eg} +iv_g\partial_x\right] \varphi^<_R(x) =\; \; \; \; \; \; \; \; \; \; \; \; \; \; \; \; \; \; \; \;  \; \; \; \; \; \; \; \; \; \; \; \; \; \; \; \; \; &\\  = V_R\left(\phi^{(ii)}_{RR}(x,0)+\phi^{(i)}_{RR}(x,0)\right) + V_L\phi^{(i)}_{RL}(x,0)/2 &+ W_R\psi^{(ii)}_{RR}(x,0)/2 + W_L\psi^{(i)}_{RL}(x,0)/2.
\end{split}\end{equation}
\begin{equation}
\begin{split}
\left[\omega - \omega_{eg} -iv_g\partial_x\right] \varphi^<_L(x) =\; \; \; \; \; \; \; \; \; \; \; \; \; \; \; \; \; \; \; \; \; \; \; \; \; \; \; \;\; \; \; \;\; \; \; \; \; \; \; \;\; \; \; \;\; \; \; \; \; \; \; \; \; \; \; \; \; \; \; \; \; &\\  = V_L\phi^{(i)}_{LL}(x,0) + V_R\left(\phi^{(iv)}_{RL}(0,x)+\phi^{(i)}_{RL}(0,x)\right)/2 + W_R\psi^{(ii)}_{LR}(x,0)/2 &+ W_L\psi^{(i)}_{LL}(x,0)/2.
\end{split}\end{equation}
In the above equations, we define the total energy of the two-photon wavepacket, $\omega = v_gk = v_g\left(k_1+k_2\right)=\omega_1+\omega_2$, where $k_1$ and $k_2$ are the wavevectors of the two photons.

The following step is to make an ansatz for the input state, i.e. the two photon wavefunction introduced through port $1$. We assume the following plane wave structure,
\begin{equation}\label{PWansatz}
\phi_{RR}^{(i)}(x_1,x_2) =A\left(e^{ik_1x_1}e^{ik_2x_2} + e^{ik_2x_1}e^{ik_1x_2}\right),
\end{equation}
which fulfills the required bosonic symmetry, and where $A$ is the normalization constant. By inserting the above ansatz into the equations, it is possible to compute the rest of the unknowns following a similar procedure as in \cite{zheng10a}. In order to express the final solutions in a more compact way, let us define the general two-photon plane wave function as
\begin{equation}
	f_{p,q} = e^{ipx_1}e^{iqx_2},
\end{equation}
where the variables may switch from $x_1,x_2$ to $x,y$ when necessary. In terms of these functions, the eigenstate coefficients normalized to $A$ are given by
\begin{equation}
\varphi_R^<(x) = 2V_R\left(\frac{e^{ik_1x}}{\omega_2-\omega_{eg}+i\gamma/2}+\frac{e^{ik_2x}}{\omega_1-\omega_{eg}+i\gamma/2}\right),
\end{equation}
\begin{equation}
\begin{split}
	\varphi_R^>(x) =& 2V_R\bigg(e^{ik_1x}\frac{t_1}{\omega_2-\omega_{eg}+i\gamma/2}+e^{ik_2x}\frac{t_2}{\omega_1-\omega_{eg}+i\gamma/2}+\\ &+\frac{2i\gamma_{dR}}{(\omega_1-\omega_{eg}+i\gamma/2)(\omega_2-\omega_{eg}+i\gamma/2)}e^{i(k-\omega_{eg})x}e^{-\gamma x/2}\bigg),
	\end{split}
\end{equation}
\begin{equation}
\varphi^<_L(x) = 2V_R\frac{r_1}{\omega_2 -\omega_{eg} +i\gamma/2}\left(e^{-ik_1x} + e^{-ik_2x}-2e^{\gamma x/2}e^{-i(k-\omega_{eg})x}\right),
\end{equation}
\begin{equation}
\phi_{RR}^{(ii)}(x_1,x_2)=
t_2f_{k_1,k_2}+t_1f_{k_2,k_1},
\end{equation}
\begin{equation}
\begin{split}
\phi_{RR}^{(iii)}& (x_1,x_2)=  t_1t_2\left(f_{k_1,k_2}+f_{k_2,k_1}\right)-\\& 2(t_1-1)(t_2-1)e^{-\gamma \vert x_2 - x_1\vert/2} \left(f_{\omega_{eg},k-\omega_{eg}}\theta(x_2-x_1) +f_{k-\omega_{eg},\omega_{eg}}\theta(x_1-x_2)\right),
\end{split}
\end{equation}
\begin{equation}
\phi_{RL}^{(ii)}(x_1,x_2)=
2\left(r_2f_{k_1,-k_2}+r_1f_{k_2,-k_1}\right),
\end{equation}
\begin{equation}
\begin{split}
\phi_{RL}^{(iv)}&(x_1,x_2) =  2\bigg(r_2t_1f_{k_1,-k_2}+r_1t_2f_{k_2,-k_1}-\\ & 2r_1(t_2-1)e^{-\gamma\vert x_1+x_2\vert/2}\left[f_{k-\omega_{eg},-\omega_{eg}}\theta(x_1+x_2)+ f_{\omega_{eg},-k+\omega_{eg}}\theta(-x_1-x_2)\right]\bigg),
\end{split}
\end{equation}
\begin{equation}
\begin{split}
	\phi_{LL}^{(i)}(x_1,x_2)=& r_1r_2\bigg(f_{-k_1,-k_2}+f_{-k_2,-k_1}- \\ & 2e^{-\gamma \vert x_2 - x_1\vert/2} \left[f_{-k+\omega_{eg},k-\omega_{eg}}\theta(x_2-x_1) +f_{-\omega_{eg},-k+\omega_{eg}}\theta(x_1-x_2)\right]\bigg),
\end{split}
\end{equation}
\begin{equation}
\psi_{RR}^{(ii)}(x,y)=
2\tilde{t}_2f_{k_1,k_2+\omega_g}+2\tilde{t}_1f_{k_2,k_1+\omega_g},
\end{equation}
\begin{equation}
\begin{split}
\psi_{RR}^{(iii)}&(x,y) =
\psi_{RR}^{(ii)}(x,y)\theta(y-x) + \\ &
2\bigg( \! \tilde{t}_2t_1f_{k_1,k_2+\omega_g}+\tilde{t}_1t_2f_{k_2,k_1+\omega_g} -2\tilde{t}_1(t_2-1)f_{k-\omega_{eg},\omega_{eg}+\omega_g}e^{\gamma(y-x)/2} \bigg)
\theta(x-y),
\end{split}
\end{equation}
\begin{equation}
\psi_{RL}^{(i)}(x,y)=
2\tilde{r}_2f_{k_1,-k_2-\omega_g}+2\tilde{r}_1f_{k_2,-k_1-\omega_g},
\end{equation}
\begin{equation}
\begin{split}
\psi_{RL}^{(iv)}&(x,y)=
\psi_{RL}^{(i)}(x,y)\theta(-x-y) + \\ &
\!2\bigg(\tilde{r}_2t_1f_{k_1,-k_2-\omega_g}+\tilde{r}_1t_2f_{k_2,-k_1-\omega_g}  -2\tilde{r}_1(t_2-1)f_{k-\omega_{eg},-\omega_{eg}-\omega_g}e^{-\gamma(y+x)/2}  
\bigg)\theta(x+y),
\end{split}
\end{equation}
\begin{equation}
\psi_{LL}^{(i)}(x,y) = 2\tilde{r}_1r_2\left(f_{-k_1,-k_2-\omega_g}+f_{-k_2,-k_1-\omega_g}-2e^{\gamma(y-x)/2}f_{-k+\omega_{eg},-\omega_{eg}-\omega_g}\right)\theta(y-x),
\end{equation}
\begin{equation}
\psi_{LR}^{(ii)}(x,y) = 2\tilde{t}_1r_2\left(f_{-k_1,k_2+\omega_g}+f_{-k_2,k_1+\omega_g}-2e^{\gamma(y+x)/2}f_{-k+\omega_{eg},\omega_{eg}+\omega_g}\right)\theta(-y-x),
\end{equation}
where $\gamma = \gamma_{dR} +\gamma_{dL}+\gamma_{uR}+\gamma_{uL},$ and the coefficients $t_j$ stand for $t(v_g k_j)$ (and the same for $r_j,\tilde{t}_j,\tilde{r}_j$).  In the above equations, whenever the energies $\gamma$, $\omega_{eg}$ or $\omega_g$ appear in the argument of exponential functions, they represent a short-hand notation for their corresponding wavevectors, $\gamma/v_g$, $\omega_{eg}/v_g$, and $\omega_g/v_g$ respectively. Finally, note that in general all the wavefunctions have two different possible components, namely a plane wave component and a two-photon \textit{bound state} proportional to $\exp(-\gamma\vert x_2 - x_1\vert)$. This nonlinear term is related to the 3LS's saturable absorber properties, and has been studied in detail in the literature \cite{fan10a}. In the plane wave limit we will work on, these bound states will not play any role in the scattering outputs. However, it is important to take into account that they could be relevant for incoming wavepackets whose frequency width is comparable to the intrinsic linewidth of the 3LS transitions \cite{zheng10a}.

 \section{Calculation of the detection probabilities.}\label{appendixC}
 
 We devote this section to the calculation of the detection probabilities $P_{mn}$, as well as to demonstrate that any contribution from the wavefunctions $\varphi_\alpha$ vanishes. We start by noticing that the two-photon components of the eigenstate calculated above split into contributions of three different types, which arise naturally from the separation between the regions $i, ii, iii, iv$ imposed by the boundary conditions. First, the state corresponding to the incoming two-photon wavepacket is given by $\phi_{RR}^{(i)}$ as defined in Eq. \ref{PWansatz}. Secondly, the two-photon wavefunctions $\phi_{RR}^{(ii)} ,\phi_{RL}^{(i)} ,\psi_{RR}^{(ii)} $, and $\psi_{RL}^{(i)}$ represent transient states, in which one of the two photons has interacted with the 3LS and the other has not. Finally, the rest of the two-photon contributions, namely $\phi_{RR}^{(iii)} ,\phi_{LL}^{(i)} ,\phi_{RL}^{(iv)} ,\psi_{RR}^{(iii)} ,\psi_{RL}^{(iv)},\psi_{LR}^{(ii)},$ and $\psi_{LL}^{(i)}$, describe the asymptotic limit in which both photons have interacted with the 3LS, and travel towards the exit ports of our system. This is a general structure for the eigenstates of any system calculated using the same method \cite{zheng10a}.

 Once the different contributions are isolated, we can properly define the so-called input state, $
 \vert \epsilon_i \rangle $, which contains only the parts of $\vert \epsilon \rangle$ for which a right-propagating photon is present in the region $x<0$. In other words, it is the fraction of the two-photon eigenstate $\vert \epsilon \rangle$ containing both the input and the transient contributions described above. From this definition, the state $\vert \epsilon_i\rangle$ can be obtained directly from the general two-photon eigenstate $\vert \epsilon \rangle$ in Eq. \ref{2photANSATZ}, by making the substitution $\phi_{RR}^{(i)} ,\phi_{RR}^{(ii)} ,\phi_{RL}^{(i)} ,\psi_{RR}^{(ii)} ,\psi_{RL}^{(i)}, \varphi_R^{<} \to 0$. 
  Finally, we can use the input state defined above to calculate the output state as $\vert \epsilon_o \rangle = \vert\epsilon\rangle - \vert \epsilon_i\rangle$. Such state contains only the asymptotic contributions to the eigenstate, as well as the functions $\varphi_R^{<} $ and $\varphi_R^{>} $ which, however, do not contribute to any detection probability as we will see below. The reason behind this definition of the output state will become clear in the following.
 
Once the output state is properly determined, we can calculate the detection probabilities $P_{mn}$. In order to obtain a general expression, we define the generalized coordinate
 \begin{equation}
 z_j = \Big\lbrace\begin{array}{lll}
x & \text{        for        } & j = 1,2 \\
y & \text{        for        } & j = 3,4, 
 \end{array}
 \end{equation}
 as well as the generalized photonic operators
  \begin{equation}
  a_i (z_i) = \left\lbrace\begin{array}{lcr}
 c_L(x)& \text{        for        } & i = 1 \\
 c_R(x)& \text{        for        } & i = 2 \\
  b_R(y)& \text{        for        } & i = 3 \\
   b_L(y)& \text{        for        } & i = 4.
  \end{array}\right.
  \end{equation}
By using these definitions we can obtain a general expression for the position probability density in ports $m$ and $n$, as
\begin{equation}\label{probDENSITIES}
\rho_{mn}(z_m,z_n) = \frac{\langle\epsilon_o\vert a_m^\dagger(z_m)a_n^\dagger(z_n)a_n(z_n)a_m(z_m)\vert\epsilon_o\rangle}{\langle\epsilon_o\vert\epsilon_o\rangle\vert_{\Gamma^*=0}}.
\end{equation}
The normalization of the above probability densities corresponds to the lossless version of the eigenstate. Otherwise,  we would be overestimating the probabilities in the lossy case, neglecting the reduction of the norm inherent to the radiative losses. This normalization method is the one implicitly chosen in all the single-photon scattering problems, both in this work and in many others \cite{fan10a,witthaut10a}. From the equation above, the total detection probability can be expressed as
\begin{equation}
P_{mn} = \frac{1}{1+\delta_{mn}}\int_{-L/2}^{L/2} dz_m\int_{-L/2}^{L/2} dz_n \rho_{mn}(z_m,z_n).
\end{equation}
In this expression, $L$ represents the length of the waveguides, which we consider infinite. Additionally, the factor $\left(1+\delta_{mn}\right)^{-1}$ prevents a double counting of the states subject to a bosonic symmetry constraint.

The first step in the calculation of the probabilities $P_{mn}$ is to prove that the norm of the output state $\vert \epsilon_o \rangle\vert_{\Gamma^*=0}$ is proportional to $L^2$, where $L \to \infty$ is the length of the waveguide. Note that this result would be trivial in the case of a bare waveguide, as it is the natural norm of a two-variable plane wave. We start by directly calculating the norm of such state as
 \begin{equation}\label{overlap}
  \begin{split}
  \langle \epsilon_o \vert \epsilon_o\rangle& = \int dx_1\int dx_2\; \left( 2\vert\phi_{LL}^{(i)}  \vert^2 + 2\vert\phi_{RR}^{(iii)} \vert^2  + \vert\phi_{RL}^{(iv)} \vert^2\right)+ \int dx \left( \vert\varphi_L^{(<)}  \vert^2+\vert\varphi_R^> \vert^2\right)+\\
  &+   \int dx\int dy\; \left(\vert\psi_{RR}^{(iii)} \vert^2+\vert\psi_{RL}^{(iv)} \vert^2  +\vert\psi_{LR}^{(ii)}  \vert^2 +\vert\psi_{LL}^{(i)}  \vert^2\right) ,
  \end{split}
  \end{equation}
  which is valid for any value of $\Gamma^*$.
 In principle, we could expand the wavefunctions by using their expressions above, but we can greatly simplify the calculation in advance. Indeed, note that apart from external factors, the overlap (\ref{overlap}) can be expressed as a sum of simple integrals, all of them with one of the following general shapes (or equivalent after a change of variables),
 \begin{equation}
  I_a = \int_0^{L/2}dx \int_{0}^{L/2}dy \;\;\; 1 = L^2/4,
  \end{equation}
 \begin{equation}
 I_b = \int_0^{L/2}dx \int_{0\text{ or }x}^{L/2}dy e^{ipx}e^{iqy} \;\;\;\;\;\;\;\;\;\;\;\;\;\;\;\;\;\;\;\; \text{where $p,q \in \mathbb{R}$,}
 \end{equation}
 \begin{equation}
  I_c = \int_0^{L/2}dx \int_{x}^{L/2}dy e^{ipx}e^{iqy} e^{-\kappa\vert y - x \vert} \;\;\;\;\;\;\;\;\;\;\;\;\;\; \text{where $p,q,\kappa \in \mathbb{R}$, and $\kappa>0$.}
  \end{equation}
 It is straightforward to demonstrate that
 \begin{equation}
 I_b \propto \delta_{pq} L^2/4,
 \end{equation}
  \begin{equation}
  I_c \propto \delta_{p0} L/2,
  \end{equation}
  which means that only the pure plane wave terms contribute to the norm, the bound states adding a negligible contribution of order $1/L \to 0$. In other words, the norm fulfills
  \begin{equation}\label{eigenNORM}
  \langle \epsilon_o \vert \epsilon_o \rangle \propto L^2 + \mathcal{O}(L)
  \end{equation}
 for any $\Gamma^*$, which is the first important result of this section.  Note that the contributions of the wavefunctions $\varphi_\alpha$ are only proportional to $L$, therefore in the limit $L\to\infty$ they do not have any weight in the norm.
 
 The two-photon detection probabilities $P_{mn}$ as defined in the main text can be split into elementary integrals exactly in the same way as we have done with the norm $\langle \epsilon_o \vert \epsilon_o \rangle$. An analogous treatment allows us to also demonstrate that
 \begin{equation}\label{PijLsquare}
  P_{mn} \propto \frac{1}{\langle \epsilon_o \vert \epsilon_o \rangle\vert_{\Gamma^*=0}} \left(L^2 + \mathcal{O}(L)\right) \to \text{constant},
  \end{equation}
 where we have made use of Eq. \ref{eigenNORM}. This apparently trivial result is extremely helpful when calculating the probabilities $P_{mn}$. Indeed, from Eq. \ref{eigenNORM} it is straightforward that the eigenstate norm will cancel out any contribution of order $\mathcal{O}(L)$ or lower, hence we only need to compute a fraction of the integrals appearing in $P_{mn}$.
 
 By using the above simplification, we can directly introduce the eigenstate wavefunctions in the definition of $P_{mn}$, obtaining the following expressions for two photons of wavevectors $k_1,k_2$ in the $L\to \infty$ limit,
 \begin{equation}
 P_{11}= R_1R_2 ,
 \end{equation}
  \begin{equation}
  P_{12}= \left(R_1T_2 + R_2 T_1\right)  ,
  \end{equation}
   \begin{equation}
   P_{22}= T_1T_2 ,
   \end{equation}
    \begin{equation}
    P_{13}= \frac{\tilde{T}_1R_2+\tilde{T}_2R_1}{2},
    \end{equation}
     \begin{equation}
     P_{14}= \frac{ \tilde{R}_1R_2+ \tilde{R}_2R_1}{2},
     \end{equation}
      \begin{equation}
      P_{23}= \frac{\tilde{T}_2\left(T_1+1\right)+\tilde{T}_1\left(T_2+1\right)}{2},
      \end{equation}
       \begin{equation}
       P_{24}= \frac{\tilde{R}_2\left(T_1+1\right)+\tilde{R}_1\left(T_2+1\right)}{2},
       \end{equation}
       \begin{equation}
       P_{33}=P_{34}=P_{44}=0.
       \end{equation}
where $\lbrace T_j,R_j,\tilde{T}_j,\tilde{R}_j\rbrace = \lbrace\vert t_j\vert^2,\vert r_j\vert^2,\vert \tilde{t}_j\vert^2,\vert \tilde{r}_j\vert^2\rbrace$. Importantly, it can be shown that in the lossless case the above probabilities add up to one,
\begin{equation}
\sum_{m=1}^4\sum_{n=m}^4 P_{mn} \Big\vert_{\Gamma^*=0}= 1.
\end{equation}
This implies that the two-photon processes whose probabilities we calculate above are the only output possibilities, and completely describe the scattering in the two-photon case. In other words, such probabilities are equivalent to the square modulus of the single-photon scattering amplitudes defined in Eqs. (\ref{t}-\ref{rtilde}). This is the reason behind the definition of the probability densities in Eq. \ref{probDENSITIES} in terms of the output state $\vert \epsilon_o \rangle$. By removing the contributions in which part or all the interaction has not yet occurred, we obtain consistent two-photon probabilities which, additionally, can be proven to coincide with the results obtained with the S-matrix formalism \cite{shi15a}. Finally, note that when we particularize the expressions of $P_{mn}$ for two equivalent photons, $k_1 = k_2$, we recover Eqs. (\ref{Pijcomplex1}-\ref{Pijcomplex2}) of the main text for $t=0$.

From the above arguments, demonstrating that the contribution of the states $\propto \varphi_{\alpha}(x) c_\alpha^\dagger(x)\vert e\rangle$ is negligible is straightforward. In principle, we could extend the definition of $P_{mn}$ and associate a detection probability to these states,
\begin{equation}\label{varp1}
P (\varphi_R^>) = \frac{1}{\langle \epsilon_o \vert \epsilon_o \rangle\vert_{\Gamma^*=0}}\int_{-L/2}^{L/2} dx \Big\vert c_R(x)\sigma_{ge} \vert\epsilon_o\rangle\Big\vert^2,
\end{equation}
\begin{equation}\label{varp3}
P (\varphi_L^<) = \frac{1}{\langle \epsilon_o \vert \epsilon_o \rangle\vert_{\Gamma^*=0}}\int_{-L/2}^{L/2} dx \Big\vert c_L(x)\sigma_{ge} \vert\epsilon_o\rangle\Big\vert^2,
\end{equation}
where $\sigma_{ge} = \vert g \rangle \langle e \vert$, and $P(\varphi_L^>)=0$ by definition as $\varphi_L^>(x)=0$ (see previous section). Now, it is straightforward to see that the largest contribution to these integrals has the form
\begin{equation}
\int_{-L/2}^{L/2}dx e^{ipx} \propto L,
\end{equation}
i.e. even the largest term in the numerator of Eqs. (\ref{varp1}-\ref{varp3}) is canceled by the denominator $\langle \epsilon_o \vert \epsilon_o \rangle \vert_{\Gamma^*=0}\propto L^2$. Any possible contribution of these states to the scattering output will then be of order $\sim 1/L \to 0$ as compared to the two-photon probabilities of Eq. (\ref{PijLsquare}). As a consequence, as we mentioned above, the detection probabilities $P_{mn}$ are the only relevant scattering variables in this case.

\section*{References}

\bibliography{Sci,books}

\providecommand{\newblock}{}
\begin{thebibliography}{10}
\expandafter\ifx\csname url\endcsname\relax
  \def\url#1{{\tt #1}}\fi
\expandafter\ifx\csname urlprefix\endcsname\relax\def\urlprefix{URL }\fi
\providecommand{\eprint}[2][]{\url{#2}}

\bibitem{obrien07a}
O'Brien J~L 2007 {\em Science\/} {\bf 318} 1567--1570

\bibitem{laucht12a}
Laucht A, P{\"u}tz S, G{\"u}nthner T, Hauke N, Saive R, Fr{\'e}d{\'e}rick S,
  Bichler M, Amann M~C, Holleitner A~W, Kaniber M and Finley J~J 2012 {\em
  Phys. Rev. X\/} {\bf 2}(1) 011014

\bibitem{lodahl15a}
Lodahl P, Mahmoodian S and Stobbe S 2015 {\em Reviews of Modern Physics\/} {\bf
  87} 347

\bibitem{vetsch10a}
Vetsch E, Reitz D, Sagu{\'e} G, Schmidt R, Dawkins S and Rauschenbeutel A 2010
  {\em Phys. Rev. Lett.\/} {\bf 104} 203603

\bibitem{goban13a}
Goban A, Hung C~L, Yu S~P, Hood J, Muniz J, Lee J, Martin M, McClung A, Choi K,
  Chang D, Painter O and Kimble H 2014 {\em Nat. Commun.\/} {\bf 5} 3808

\bibitem{kimble08a}
Kimble H 2008 {\em Nature\/} {\bf 453} 1023--1030

\bibitem{mitsch14a}
Mitsch R, Sayrin C, Albrecht B, Schneeweiss P and Rauschenbeutel A 2014 {\em
  Nature Commun.\/} {\bf 5} 5713

\bibitem{petersen14a}
Petersen J, Volz J and Rauschenbeutel A 2014 {\em Science\/} {\bf 346} 67--71

\bibitem{coles15a}
Coles R~J, Price D~M, Dixon J~E, Royall B, Clarke A~M, Fox P, Skolnick M~S and
  Makhonin M~N {\em arXiv\/}  1506.02266

\bibitem{sollner15a}
S\"ollner I, Mahmoodian S, Hansen S~L, Midolo L, Javadi A, Pregnolato T,
  El-Ella H, Lee E~H, Song J~D, Stobbe S and Lodahl P 2015 {\em Nat Nano\/}
  {\bf 10} 775--778

\bibitem{lodahl16}
 2016 {\em arXiv\/}  1608.00446

\bibitem{chang07b}
Chang D~E, S{\o}rensen A~S, A D~E and Lukin M~D 2007 {\em Nature Physics\/}
  {\bf 3}

\bibitem{zheng13a}
Zheng H and Baranger H~U 2013 {\em Phys. Rev. Lett.\/} {\bf 110}(11) 113601

\bibitem{ralph15a}
Ralph T~C, S\"ollner I, Mahmoodian S, White A~G and Lodahl P 2015 {\em Phys.
  Rev. Lett.\/} {\bf 114}(17) 173603

\bibitem{dzsotjan10a}
Dzsotjan D, S{\o}rensen A~S and Fleischhauer M 2010 {\em Phys. Rev. B\/} {\bf
  82}(7) 075427

\bibitem{gonzaleztudela11a}
Gonzalez-Tudela A, Martin-Cano D, Moreno E, Martin-Moreno L, Tejedor C and
  Garcia-Vidal F~J 2011 {\em Phys. Rev. Lett.\/} {\bf 106} 020501

\bibitem{ramos14a}
Ramos T, Pichler H, Daley A~J and Zoller P 2014 {\em Phys. Rev. Lett.\/} {\bf
  113}(23) 237203

\bibitem{pichler15a}
Pichler H, Ramos T, Daley A~J and Zoller P 2015 {\em Phys. Rev. A\/} {\bf
  91}(4) 042116

\bibitem{gonzalezballestero15a}
Gonzalez-Ballestero C, Gonzalez-Tudela A, Garcia-Vidal F~J and Moreno E 2015
  {\em Phys. Rev. B\/} {\bf 92}(15) 155304

\bibitem{paulisch16a}
Paulisch V, Kimble H and Gonz{\'a}lez-Tudela A 2016 {\em New Journal of
  Physics\/} {\bf 18} 043041

\bibitem{gonzaleztudela15a}
Gonz\'alez-Tudela A, Paulisch V, Chang D~E, Kimble H~J and Cirac J~I 2015 {\em
  Phys. Rev. Lett.\/} {\bf 115}(16) 163603

\bibitem{gonzaleztudela16a}
Gonz{\'a}lez-Tudela A, Paulisch V, Kimble H and Cirac J 2016 {\em
  arXiv:1603.01243\/}

\bibitem{jalas13a}
Jalas D, Petrov A, Eich M, Freude W, Fan S, Yu Z, Baets R, Popovic M, Melloni
  A, Joannopoulos J~D, Vanwolleghem M, Doerr C~R and Renner H 2013 {\em Nat
  Photon\/} {\bf 7} 579--582

\bibitem{huang15a}
Huang Y, Veronis G and Min C 2015 {\em Opt. Express\/} {\bf 23} 29882--29895

\bibitem{weibin15a}
Yan W~B, Liu B, Zhou L and Fan H 2015 {\em EPL (Europhysics Letters)\/} {\bf
  111} 64005 \urlprefix\url{http://stacks.iop.org/0295-5075/111/i=6/a=64005}

\bibitem{mascarenhas14a}
Mascarenhas E, Gerace D, Valente D, Montangero S, Auffeves A and Santos M~F
  2014 {\em EPL (Europhysics Letters)\/} {\bf 106} 54003

\bibitem{shen16a}
Shen Z, Zhang Y~L, Chen Y, Zou C~L, Xiao Y~F, Zou X~B, Sun F~W, Guo G~C and
  Dong C~H 2016 {\em arXiv\/}  1604.02297

\bibitem{hafezi12a}
Hafezi M and Rabl P 2012 {\em Opt. Express\/} {\bf 20} 7672--7684

\bibitem{rosenblum16a}
Rosenblum S, Bechler O, Shomroni I, Lovsky Y, Guendelman G and Dayan B 2016
  {\em Nat Photon\/} {\bf 10} 19--22

\bibitem{fratini16a}
Fratini F and Ghobadi R 2016 {\em Phys. Rev. A\/} {\bf 93}(2) 023818

\bibitem{dai15a}
Dai J, Roulet A, Le H~N and Scarani V 2015 {\em Phys. Rev. A\/} {\bf 92}(6)
  063848 \urlprefix\url{http://link.aps.org/doi/10.1103/PhysRevA.92.063848}

\bibitem{mascarenhas16a}
Mascarenhas E, Santos M~F, Auff\`eves A and Gerace D 2016 {\em Phys. Rev. A\/}
  {\bf 93}(4) 043821

\bibitem{chen15a}
Chen X~Y, Zhang F~Y and Li C 2015 {\em arXiv\/}  1512.04154

\bibitem{fratini14a}
Fratini F, Mascarenhas E, Safari L, Poizat J~P, Valente D, Auff\`eves A, Gerace
  D and Santos M~F 2014 {\em Phys. Rev. Lett.\/} {\bf 113}(24) 243601

\bibitem{yuan15a}
Yuan L, Xu S and Fan S 2015 {\em Opt. Lett.\/} {\bf 40} 5140--5143

\bibitem{sayrin15a}
Sayrin C, Junge C, Mitsch R, Albrecht B, O'Shea D, Schneeweiss P, Volz J and
  Rauschenbeutel A 2015 {\em Phys. Rev. X\/} {\bf 5}(4) 041036

\bibitem{fan10a}
Fan S, Kocaba\ifmmode~\mbox{\c{s}}\else \c{s}\fi{} S~E and Shen J~T 2010 {\em
  Phys. Rev. A\/} {\bf 82}(6) 063821
  \urlprefix\url{http://link.aps.org/doi/10.1103/PhysRevA.82.063821}

\bibitem{witthaut10a}
Witthaut D and S{\o}rensen A~S 2010 {\em New Journal of Physics\/} {\bf 12}
  043052

\bibitem{bamba11a}
Bamba M, Imamo\ifmmode~\breve{g}\else \u{g}\fi{}lu A, Carusotto I and Ciuti C
  2011 {\em Phys. Rev. A\/} {\bf 83}(2) 021802

\bibitem{majumdar12a}
Majumdar A, Bajcsy M, Rundquist A and Vu{\v{c}}kovi{\'c} J 2012 {\em Physical
  review letters\/} {\bf 108} 183601

\bibitem{shi13a}
Shi T and Fan S 2013 {\em Phys. Rev. A\/} {\bf 87}(6) 063818

\bibitem{koshino09a}
Koshino K 2009 {\em Physical Review A\/} {\bf 79} 013804

\bibitem{chang15a}
Chang Y, Gonz{\'a}lez-Tudela A, S{\'a}nchez-Mu{\~n}oz C, Navarrete-Benlloch C
  and Shi T 2015 {\em rXiv:1510.07307\/}

\bibitem{sanchezburillo16a}
S{\'a}nchez-Burillo E, Mart{\'\i}n-Moreno L, Garc{\'\i}a-Ripoll J and Zueco D
  2016 {\em arXiv:1602.05603\/}

\bibitem{kolchin15a}
Kolchin P, Pholchai N, Mikkelsen M~H, Oh J, Ota S, Islam M~S, Yin X and Zhang X
  2015 {\em Nano Letters\/} {\bf 15} 464--468 pMID: 25432015

\bibitem{shi09a}
Shi T and Sun C~P 2009 {\em Phys. Rev. B\/} {\bf 79}(20) 205111

\bibitem{shi15a}
Shi T, Chang D~E and Cirac J~I 2015 {\em Phys. Rev. A\/} {\bf 92}(5) 053834

\bibitem{caneva15a}
Caneva T, Manzoni M~T, Shi T, Douglas J~S, Cirac J~I and Chang D~E {\em
  arXiv:1501.04427\/}

\bibitem{xu15a}
Xu S and Fan S {\em arXiv:1502.06049\/}

\bibitem{zheng10a}
Zheng H, Gauthier D~J and Baranger H~U 2010 {\em Phys. Rev. A\/} {\bf 82}(6)
  063816

\bibitem{gonzalezballestero13a}
Gonzalez-Ballestero C, Garci�a-Vidal F~J and Moreno E 2013 {\em New Journal
  of Physics\/} {\bf 15} 073015
  \urlprefix\url{http://stacks.iop.org/1367-2630/15/i=7/a=073015}

\end{thebibliography}

\end{document}